\documentclass[12pt]{article}
\usepackage{amsmath,amssymb,bm,bbm,mathrsfs,amscd}
\usepackage{calc,cite}
\usepackage{geometry}
\usepackage{color}
\usepackage{amsthm,placeins}
\usepackage{pdfpages,xurl}
\usepackage{hyperref}
\usepackage{graphicx}
\usepackage{dcolumn}
\usepackage{bm}
\renewcommand{\eqref}[1]{Eq.~(\ref{#1})}  

\usepackage{pgfplots,subfig}
\def\l {3} 
\def\h {1.8} 
\def\lf {5.5} 
\def\hf {3} 
\def\lNO {5.5} 
\def\hNO {3} 
\usepackage{subfig,tikz,pgfplots,xurl}
\pgfplotsset{compat=newest, yticklabel style={
        /pgf/number format/fixed},
scaled y ticks=false,
colormap={CM}{rgb255(0cm)=(255,255,255);rgb255(2.5cm)=(156,156,255);rgb255(5cm)=(0,0,255)}}

\definecolor{mygreen}{RGB}{72,160,66}
\definecolor{myblue}{RGB}{20,155,204}
\definecolor{myyellow}{RGB}{190,170,0}
\definecolor{myviolet}{RGB}{195,8,255}
\definecolor{myred}{RGB}{204,0,0}
\definecolor{myblue}{RGB}{0,66,255}

\title{Game-theoretic modeling of collective decision-making during epidemics}
\author{Mengbin Ye$^{1}$, Lorenzo Zino$^{2}$, Alessandro Rizzo$^{3,4}$, Ming Cao$^2$}

\date{\normalsize $^1$Optus--Curtin Centre of Excellence in Artificial Intelligence, Curtin University, Perth, Australia\\
$^2$Faculty of Science and Engineering, University of Groningen, 9747 AG Groningen, Netherlands\\
$^3$Department of Electronics and Telecommunications, Politecnico di Torino, 10129 Turin, Italy\\
$^4$Office of Innovation, New York University Tandon School of Engineering, Brooklyn NY 11201, USA\\\quad\\
Correspondence should be addressed to: \url{mengbin.ye@curtin@edu.au}}

\begin{document}

\maketitle
\newpage

\begin{abstract}

The spreading dynamics of an epidemic and the collective behavioral pattern of the population over which it spreads are deeply intertwined and the latter can critically shape the outcome of the former. Motivated by this, we design a parsimonious game-theoretic behavioral--epidemic model, in which an interplay of realistic factors shapes the co-evolution of individual decision-making and epidemics on a network. Although such a co-evolution is deeply intertwined in the real-world, existing models schematize population behavior as instantaneously reactive, thus being unable to capture human behavior in the long term. Our model offers a unified framework to model and predict complex emergent phenomena, including successful collective responses, periodic oscillations, and resurgent epidemic outbreaks. The framework also allows to assess the effectiveness of different policy interventions on ensuring a collective response that successfully eradicates the outbreak. Two case studies, inspired by real-world diseases, are presented to illustrate the potentialities of the proposed model.

\end{abstract}

\maketitle

\section*{Introduction}

The collective adoption of appropriate behavior by a population is crucial to respond to an epidemic, especially when pharmaceutical interventions are absent or logistical challenges prevent their widespread deployment~\cite{Bavel2020,Bedson2021}. However, classical mathematical epidemic models often consider oversimplified behavioral response~\cite{Pastor-Satorras2015}. To fill in this gap, awareness-based models have been proposed~\cite{Funk2010,Perra2011,Sahneh2012,Granell2013,RizzoPRE2014,Wang2015,Verelst2016,Weitz2020,Gozzi2021}, in which the epidemic process co-evolves with the spread of the awareness of the outbreak. While these models have demonstrated effectiveness in capturing the early-stage, immediate behavioral response to an epidemic, they are limited because they assume  fully rational, purely instantaneous, and reactive decision-making in the population. Such models therefore fail to capture the very range of factors that affect real-world behavioral responses over the whole course of an epidemic, such as social influence~\cite{Tuncgenc2021influence}, perceived infection risk~\cite{Poletti2010}, accumulating fatigue and socio-economic costs~\cite{nicola2020socio,pedro2020},
bounded rationality in individuals' decisions~\cite{Simon2000}, and the impact of government-mandated  interventions~\cite{Flaxman2020,Perra2021}.

The world is not new to epidemics that evolve over long time horizons of several months or even years, persisting until effective drugs and vaccines are developed and then  made widely available~\cite{Piret2021}, making purely reactive models of limited efficacy. This calls for a paradigm shift in mathematical modeling, from reactive and fully rational behavioral responses~\cite{Funk2010,Perra2011,Sahneh2012,Granell2013,Wang2015,Verelst2016}, to a long-term outlook where complex behavioral dynamics arise at the individual-level and co-evolve at the same time scale of the epidemic spreading. Game-theoretic models have proved to be effective  to reproduce similar complex decision-making mechanisms, thereby capturing  realistic behavioral responses in several fields~\cite{handbook_game}.

Here, we propose a game-theoretic model that is specifically designed to account for long-term and bounded-rational decision-making, and show it is able to reproduce the complex and concurrent evolution of behavioral response and epidemic spreading that is well-known and documented in the real-world~\cite{Bavel2020}. Several efforts have been proposed across similar research avenues, although with different and narrower angles. For example, imitation mechanisms~\cite{Poletti2010,pedro2020,kabir2020,wei2020game} rely on a population-level modeling that can capture only limited features of such complex behavioral dynamics. Recently, an individual-level imitation-driven mechanism that accounts for the perceived risk of infection and immediate costs for adopting protective behaviors has been proposed and analyzed, showing that it may generate sustained steady oscillations~\cite{Just2017,Steinegger2020}. Following a different approach,  game-theoretic modeling of vaccination adoption have been proposed~\cite{Bauch2004,Fu2010,Zhang2014,Zhang2017vaccination,chen2019vaccine,Chang2020,wells2020}. These models rely on a time-scale separation between the epidemic spreading and the behavioral decision, which is typically made just at the beginning of each epidemic season. However, such a time-scale separation does not capture general behavioral response. 

Here, we adopt a network approach with individual granularity~\cite{Pastor-Satorras2015} and a co-evolution of the two processes, at the same time scale, under the impact of the entire range of factors discussed in the above (i.e., including perceived risk, immediate costs,  accumulating fatigue,  social influence, and bounded rationality). Hence, our model is designed to be able to capture the individual-level responses and time-varying contagion patterns, whose \emph{co-evolution} collectively shapes the epidemic outbreak. Our approach enables the \textit{explicit and concurrent} inclusion of the most salient factors that each individual  trades off when deciding their time-varying behavioral response to an ongoing epidemic. The central contribution of our work is the design of a unified and parsimonious mathematical framework for the co-evolution of the decision-making and the epidemic outbreak, which can be coupled with any existing compartmental model~\cite{brauer2011mathematical}, and thus it can be tailored to study the key features of any real-world disease. As we shall illustrate through some simple case studies, the proposed framework is able to capture and reproduce complex realistic behavioral response by the population, including successful collective responses leading to the eradication of the disease, periodic oscillations, and weak responses leading to the emergence of multiple epidemic waves.

\begin{figure}
    \centering
\includegraphics[width=\linewidth]{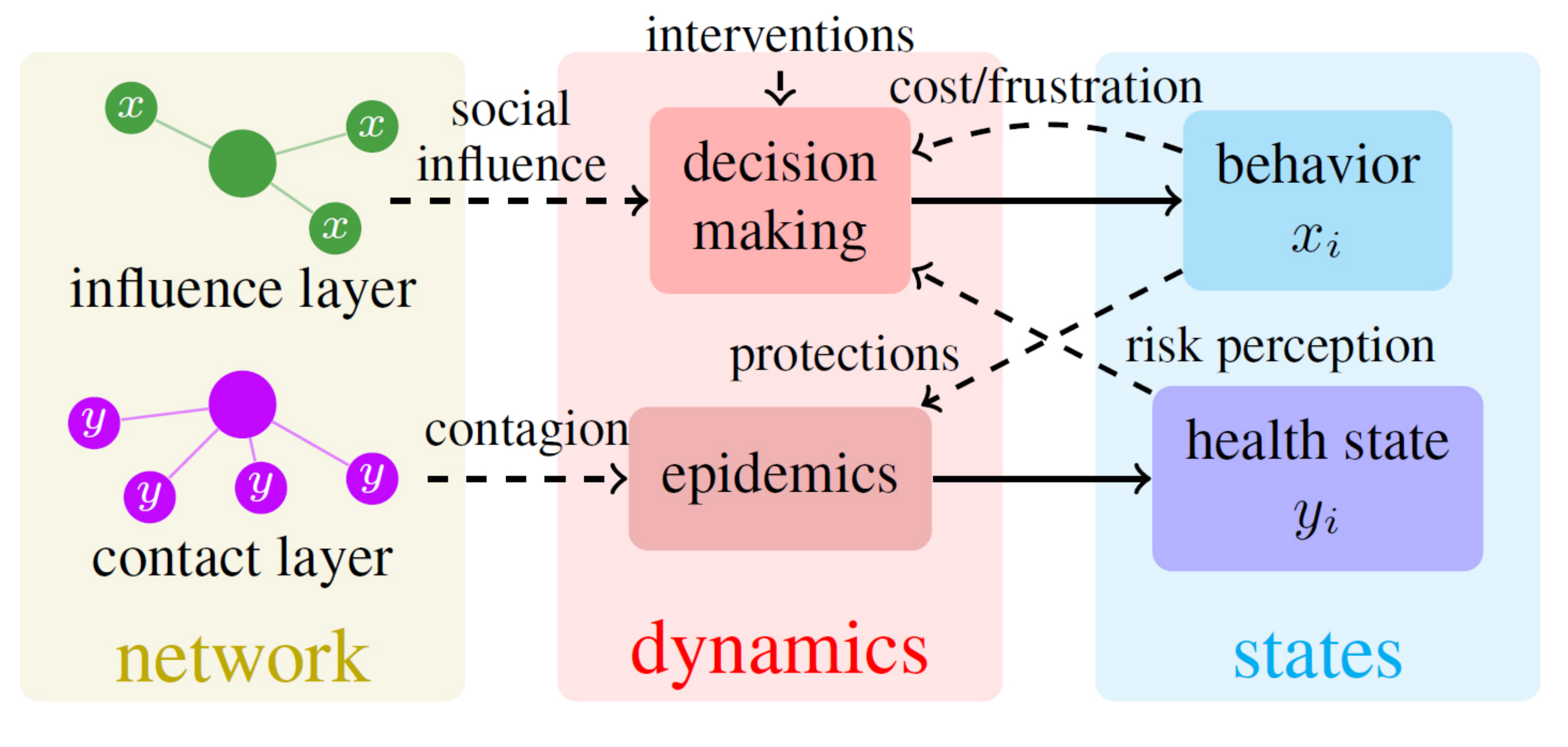}
\caption{Schematic of the co-evolutionary paradigm. }
    \label{fig:schematic}
\end{figure}

\section*{Model}

We consider a population $\mathcal V$ of $n$ individuals. Each individual $i\in \mathcal V$ is characterized by a two-dimensional variable $(x_i(t),y_i(t))$, which models their \emph{social behavior} and \emph{health state} at the discrete time $t\in\mathbb Z_{\geq 0}$, respectively. The social behavior of individual $i$ is captured by the binary variable $x_i(t)\in\{0,1\}$, which expresses whether $i$ adopts self-protective behaviors ($x_i(t)=1$), such as physical distancing~\cite{Bavel2020}, or discards this opportunity ($x_i(t)=0$). The health state $y_i(t)$ takes values in a discrete set of compartments $\mathcal A$. For example, $\mathcal A = \{ S, I \}$ is selected to model the  susceptible--infected--susceptible (SIS) epidemic process exemplified in this work~\cite{brauer2011mathematical}. A global observable $z(t)$ quantifies the \emph{detectable prevalence} of the epidemic at time $t$: $z(t):=\frac1n\left|\{i:y_i(t) = I\}\right|$, where $| \cdot |$ denotes a set's cardinality.  The paradigm is amenable to extensions to capture different levels of protection through the selection of a different support for $x_i(t)$, while further compartments added to $\mathcal{A}$ can capture additional features of the epidemic process~\cite{brauer2011mathematical}. 

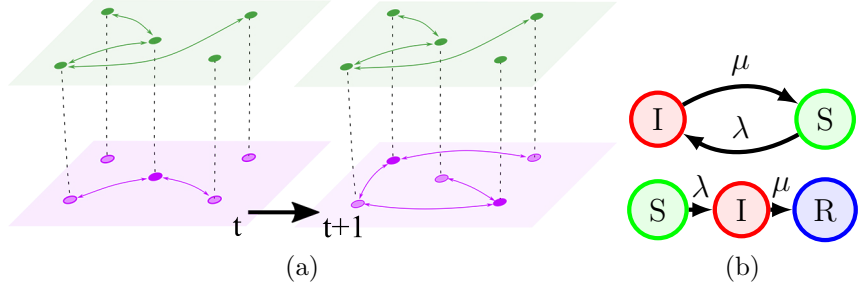
\begin{figure}
    \centering
        \subfloat[]{
    \def\svgwidth{.6\linewidth}
    \includegraphics{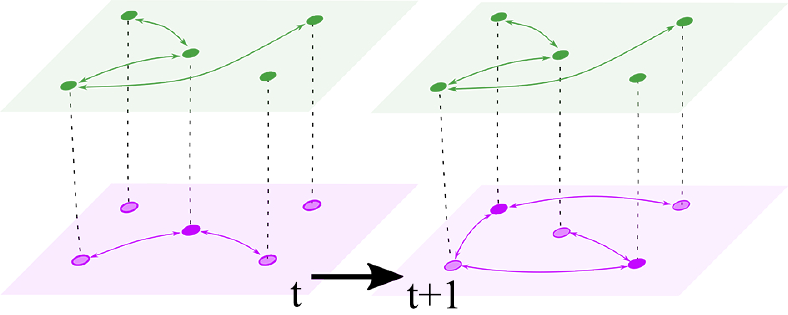}\label{fig:network}}\subfloat[]{ \begin{tikzpicture}
\node[draw=red, fill=red!10,circle, ultra thick,,minimum size=0.6cm] (I1) at (0.2,1.2) { I};
\node[draw=green, fill=green!10,circle, ultra thick,,minimum size=0.6cm]  (S1) at (2.4,1.2) { S};
\path [->,>=latex,ultra thick]  (S1) edge[bend left =30]   node [above] {{$\lambda$}} (I1);
\path [->,>=latex,ultra thick]  (I1) edge[bend left =30]   node [above] {{$\mu$}} (S1);

\node[draw=green, fill=green!10,circle, ultra thick,minimum size=0.6cm]  (S2) at (0.2,0) { S};
\node[draw=red, fill=red!10,circle, ultra thick,minimum size=0.6cm] (I2) at (1.3,0) { I};
\node[draw=blue, fill=blue!10,circle, ultra thick,minimum size=0.6cm] (R2) at (2.4,0) { R};
\path [->,>=latex,ultra thick]  (S2) edge   node [above] {{$\lambda$}} (I2);
\path [->,>=latex,ultra thick]  (I2) edge   node [above] {{$\mu$}} (R2);
\end{tikzpicture} \label{fig:schematic_models}}
    \caption{Illustration of the network model and the epidemic progression. In (a), two time-steps in the two-layer network representation. The upper layer (green) shows the static influences, the lower layer (violet) time-varying physical contacts. In (b), state transitions of the SIS (above) and SIR models (below). }
    \label{fig:model}
\end{figure}

The decision-making and disease spreading in the population co-evolve, mutually influencing each other on a two-layered network $\mathcal G=({\mathcal V,\mathcal E_{I},\mathcal E_C(t)})$~\cite{Boccaletti2014}, as schematized in Fig.~\ref{fig:schematic}. The set of undirected links $\mathcal E_{I}$ defines the static \textit{influence layer}, capturing \emph{social influence} between individuals in their decision-making processes. The \textit{contact layer} is defined through a time-varying set of undirected links $\mathcal E_{C}(t)$, which represent the \emph{physical contacts} between pairs of individuals that are the avenues for the transmission of the disease. The temporal formation mechanism of the contact layer, illustrated in Fig.~\ref{fig:network}, is  general and can be generated according to any model of time-varying networks~\cite{Holme2012,Perra2012,Holme2015,Valdano2015,Zino2016,Koher2019}. 

At each discrete time-step $t$, every individual $i$ enacts a decision-making process on the adoption of self-protective behaviors, according to an evolutionary game-based mechanism termed logit learning~\cite{blume1995best_response}. We define two \emph{payoff} functions $\pi_i^0(t)$ and $\pi_i^1(t)$, which represent a combination of socio-psychological, economic, and personal benefits received by individual $i$ for enacting behaviors $x_i = 0$ and $x_i = 1$ at time $t$, respectively. This individual then adopts self-protective behaviors  with a probability equal to
\begin{equation}\label{eq:log-linear}
    \mathbb P[x_i(t+1)=1]=\frac{\exp\{\beta \pi_i^1(t)\}}{\exp\{\beta \pi_i^0(t)\}+\exp\{\beta \pi_i^1(t)\}}\,;
\end{equation}
otherwise the individual will adopt $x_i(t+1)=0$. The parameter $\beta\in[0,\infty)$ measures an individual's \emph{rationality} in the decision-making process. We have assumed for simplicity that  $\beta$ is homogeneous among all individuals, but this is easily generalizable to a heterogeneous $\beta_i$ distribution. Notice that if $\beta=0$, individuals make decisions uniformly at random, while for $\beta\to \infty$, individuals apply perfect rationality to select the behavior with highest payoff. This best-response behavior is myopic, i.e., individuals do not look forward in time to optimize a sequence of decisions. Myopic behavior is reasonable given the uncertain nature of a long-lasting epidemic.

Payoffs are defined as
\begin{subequations}
\begin{align}\label{eq:pi0}
    \pi_i^0(t)\,:=\,&\frac{1}{d_i}\sum_{j:(i,j)\in\mathcal E_I}\big(1-x_j(t)\big)-u(t),\\
    \label{eq:pi1}
    \pi_i^1(t)\,:=\,&\frac{1}{d_i}\sum_{j:(i,j)\in\mathcal E_I}x_j(t)+r\big(z(t)\big)-f_i(t),
\end{align}
\end{subequations}
where $d_i:=|\{j:(i,j)\in\mathcal E_I\}|$ is the degree of node $i$ on the influence layer, and contain the following four terms, directly related to behavioral and social  factors that shape the epidemic dynamics. \medskip

{\bf Social influence.} The first term in Eqs.~(\ref{eq:pi0})--(\ref{eq:pi1}), inspired by network coordination games~\cite{Jackson2015}, captures the \emph{social influence} of neighboring individuals and the individual's desire to coordinate with them on the behavioral response~\cite{Tuncgenc2021influence}. The role of social influence should be understood by viewing the first terms of \eqref{eq:pi0} and \eqref{eq:pi1} together; as more of the neighbors of individual $i\in \mathcal{V}$ adopt self-protection or do not adopt self-protection, then individual $i$ also has more incentive to adopt or not adopt, respectively. This ensures that individuals tend to conform and coordinate with one another; coordination and conformity are prevalent factors for many different human behaviors, including social norms and conventions~\cite{cialdini2004social_conformity,peytonyoung2015social_norms,young1993evolution}, diffusion of social innovation~\cite{montanari2010spread_innovation,young2009innovation}, and also in individuals' decisions concerning the \textit{behavioral response} to epidemics~\cite{Tuncgenc2021influence}.
\medskip

{\bf Policy interventions.} The time-varying term $u(t)\geq 0$ in~\eqref{eq:pi0} represents the impact of \emph{nonpharmaceutical interventions} enforced by public authorities to discourage dangerous behaviors, e.g., lockdowns, see~\cite{Perra2021} for more details.\medskip

{\bf Risk perception.} The \emph{risk-perception function} $r(z) : [0, 1] \to \mathbb{R}_{\geq 0}$ in~\eqref{eq:pi1} is a monotonically nondecreasing function of the detectable prevalence $z$, which models the population's reaction to the spread of the disease. This function is amenable to several generalizations, e.g., to capture imperfect or delayed information due to real-world testing logistics, or heterogeneity of the function in the population. In its simplest formulation (which is adopted in the  case studies presented in this paper), it can be assumed to be a power function $r(z)=kz^\alpha$, with $k>0$ as a scaling factor and $\alpha>0$ that determines the characteristics of the population. Specifically, $\alpha\in(0,1)$ models cautious populations, where a small initial outbreak causes a large increase in the risk perception; $\alpha=1$ captures a population whose reaction grows linearly with the epidemic prevalence observed; and $\alpha>1$ captures populations that underestimate the risk, and the epidemic prevalence must be large before the risk perception plays an important role in the decision-making process. \medskip

{\bf Cost of self-protective behavior.} The negative impact of adopting self-protections is captured in~\eqref{eq:pi1} by the \emph{frustration function} 
\begin{equation}
    f_i(t)=c+\sum_{s=1}^t\gamma^s cx_i(t-s),
\end{equation}
where $c\geq 0$ quantifies the social, psychological, and economic \emph{immediate cost} per unit-time, e.g., related to the inability to socialize, work from the office, enjoy public spaces, etc., 
and $\gamma\in[0,1]$ is the \emph{accumulation factor}~\cite{Qiu2020,pedro2020}. When $\gamma=0$,  an individual accounts only for the immediate cost, and as $\gamma$ increases, the impact of all past decisions on the payoff increases.  This may reflect the accumulating nature of fatigue, stress, and economic losses~\cite{nicola2020socio,pedro2020,Qiu2020}. Thus, $f_i(t)\geq 0$ reflects accumulative costs for individual $i$ up to time $t$. Note that the function $f_i(t)$ could be extended by adding further features of the frustration mechanism, including nonlinearities. \medskip

To summarize, the payoff that an individual $i\in\mathcal V$ would receive for not adopting self-protective behavior in \eqref{eq:pi0} is equal to the difference of two terms. Besides the social influence term, the other term reduces the payoff to represent the implementation of policies to disincentivize nonprotective behaviors; the effectiveness of policies in shaping behaviors has been extensively analyzed in the recent literature~\cite{Flaxman2020,Perra2021}. The payoff for adopting self-protective behavior in \eqref{eq:pi1}, instead, is equal to the sum of two positive contributions, and decreased by a third one. The first term accounts for the social influence. The second term, associated with the risk perception, captures an increased incentive to adopt self-protective behavior due to the endogenous fear of increased risk to infection as the disease spreads (and conversely, a lower risk is perceived as the epidemic dwindles due to the growing optimism of returning to normal). This term has some analogies to the mechanism of purely reactive awareness-based models, but which do not consider the other factors detailed above~\cite{Funk2010,Perra2011,Sahneh2012,Granell2013,RizzoPRE2014,Wang2015,Verelst2016,Weitz2020,Gozzi2021}. A similar implementation of this term is often present in the payoffs of imitation-based game-theoretic models~\cite{Poletti2010,pedro2020,kabir2020,wei2020game,Steinegger2020}. Finally, the last term reduces the payoff to account for the immediate and accumulated social, psychological, and economic costs associated with the adoption of self-protective behavior~\cite{Qiu2020,nicola2020socio}. A similar term ---without the accumulation mechanism--- has been considered in some imitation-based models~\cite{Steinegger2020}. 

Concurrently with the behavioral decision, at each time-step $t$, every individual $i$ that does not adopt self-protections and is susceptible (i.e., $x_i(t)=0$ and $y_i(t)=S$) may become infected upon contact with an infected individual $j: y_j(t)=I$, with a per-contact \emph{infection probability} $\lambda\in[0,1]$. We introduce a parameter $\sigma\in[0,1]$ that represents the \emph{effectiveness} of self-protective behavior in preventing contagion, and we assume that the adoption of self-protection, $x_i(t)=1$, does not affect the individual's probability of transmitting the disease. 
Hence, considering an SIS epidemic model (see Fig.~\ref{fig:schematic_models}), the contagion probability for  individual $i\in\mathcal V$ evolves in time as  
\begin{equation}\label{eq:sis}
    \mathbb P[y_i(t+1)=I|y_i(t)=S]=(1-\sigma x_i(t))\Big(1-(1-\lambda)^{N_i(t)}\Big),
\end{equation}where \begin{equation}
    N_i(t):=|\{j\in\mathcal V:(i,j)\in\mathcal E_C(t),\,y_j(t)=I\}|
\end{equation} is the number of infectious physical contacts of node $i$ at time $t$. Note that the model can be extended to account for mutual protection by adding an additional parameter and expanding the term $N_i(t)$, depending on the behavior of the neighbors. Besides the contagion, at each time-step $t$, every infected individual $i$ recovers with probability $\mu \in (0, 1]$, becoming susceptible again to the disease, i.e., \begin{equation}\mathbb P[y_i(t+1)=S\,|\,y_i(t)=I]=\mu.\end{equation} Table~\ref{tab:notation} summarizes the notation. 

\begin{table}
\centering
\caption{Notation.}\label{tab:notation}
\begin{tabular}{r| l r|l}
$n$& population size&$\lambda$&infection probability\\
$\mathcal E_I$& influence layer edges&$\mu$&recovery probability\\
$\mathcal E_C(t)$& contact layer edges&$\beta$   &   rationality\\
$N_i(t)$& infectious contacts of $i$&$u(t)$   &   policy interventions\\
$\sigma$& efficacy of self-protections&$r(z)$   &   risk perception function\\
$x_i(t)$   &   behavior of $i$&$c$   &   immediate cost\\
$y_i(t)$   &   health state of $i$&$\gamma$   &   accumulation factor\\
$z(t)$   &   detected prevalence&$\quad f_i(t)$   &   frustration function of $i$\vspace{-12pt}
\end{tabular}
\end{table}

\section*{Results}

\begin{figure*}
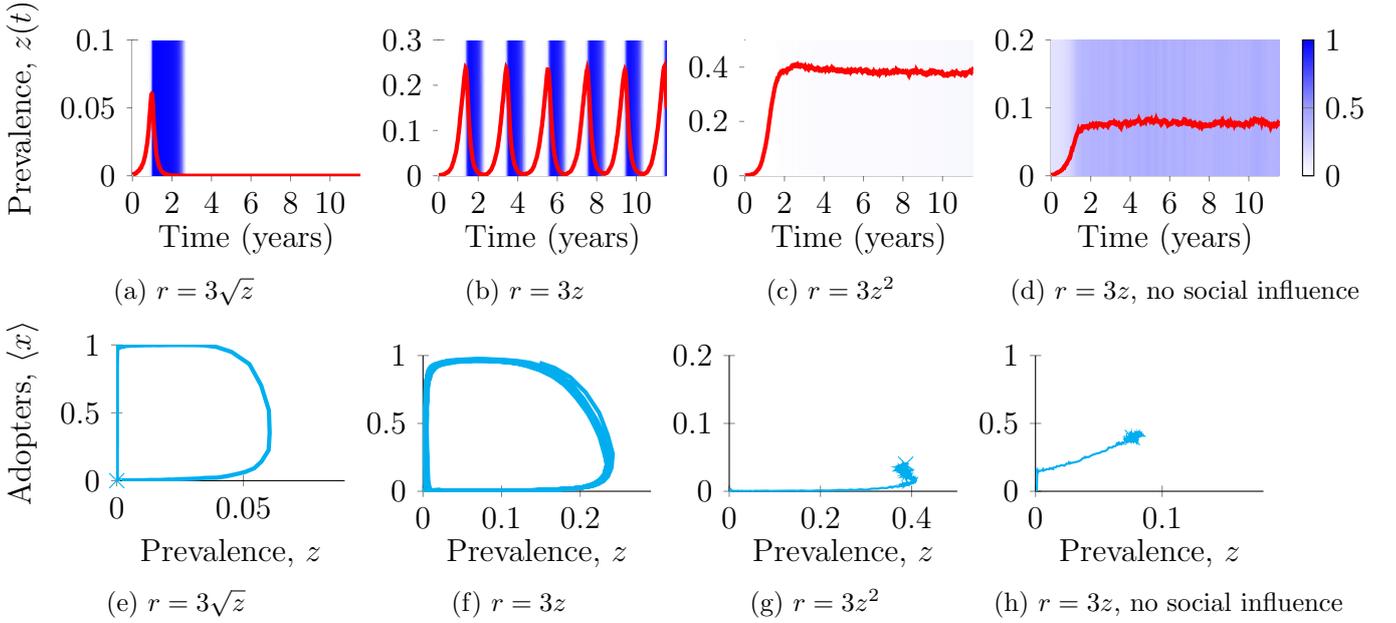

      \subfloat[$r=3\sqrt z$]{\input{sis1.tex}\label{fig:sis1}}
      \subfloat[$r=3 z$]{\input{sis2.tex}\label{fig:sis2}}
      \subfloat[$r=3 z^2$]{\input{sis3.tex}\label{fig:sis3}}
      \subfloat[$r=3 z$, no social influence]{\input{sis4.tex}\label{fig:sis4}}\\\vspace{-.4cm}
      
      \subfloat[$r=3\sqrt z$]{\begin{tikzpicture}
\begin{axis}[%
axis lines=left,
point meta min=0, 
   point meta max=1,
x    axis line style={-},
y    axis line style={-},
width=\l cm,
height=\h cm,
scale only axis,
xmin=0,
xmax=0.09,
ymin=0,
xtick distance=.05,
ytick distance=.5,
ymax=1,
ylabel={Adopters, $\langle x\rangle $},
xlabel={Prevalence, $z$},
xticklabel style={
        /pgf/number format/fixed,
        /pgf/number format/precision=5
},
scaled x ticks=false,
]

\addplot [color=cyan,ultra thick]
  table[row sep=crcr]{%
0.001	0\\
0.00115	0.00095\\
0.00115	0.001\\
0.00125	0.00125\\
0.0014	0.00075\\
0.00135	0.00095\\
0.00135	0.00135\\
0.0014	0.00105\\
0.00155	0.0012\\
0.00165	0.0012\\
0.00185	0.0016\\
0.0025	0.00085\\
0.00275	0.00085\\
0.00295	0.00125\\
0.00325	0.00165\\
0.0033	0.00215\\
0.0035	0.0018\\
0.0037	0.0018\\
0.00425	0.00175\\
0.00475	0.00235\\
0.0053	0.00235\\
0.0056	0.002\\
0.00555	0.00245\\
0.00585	0.0027\\
0.00655	0.00265\\
0.00755	0.00255\\
0.0082	0.00235\\
0.0085	0.0031\\
0.0092	0.0031\\
0.0098	0.00265\\
0.01105	0.00325\\
0.0122	0.0041\\
0.01325	0.005\\
0.01405	0.0051\\
0.01575	0.0051\\
0.0168	0.00465\\
0.01895	0.00665\\
0.02065	0.0083\\
0.0225	0.0079\\
0.024	0.01115\\
0.0268	0.0122\\
0.0296	0.014\\
0.03245	0.01595\\
0.03455	0.0201\\
0.0396	0.0224\\
0.04315	0.03225\\
0.04645	0.045\\
0.05065	0.06385\\
0.0542	0.09095\\
0.0571	0.1401\\
0.05995	0.2248\\
0.06035	0.35115\\
0.06005	0.51845\\
0.057	0.7015\\
0.05235	0.85895\\
0.04535	0.948\\
0.0394	0.9848\\
0.03455	0.99635\\
0.0316	0.99835\\
0.0277	0.9989\\
0.02505	0.9991\\
0.02145	0.99895\\
0.01855	0.99825\\
0.01625	0.9977\\
0.0143	0.9985\\
0.01305	0.99735\\
0.01175	0.9979\\
0.0104	0.9969\\
0.0099	0.99645\\
0.0084	0.9961\\
0.0071	0.99525\\
0.00645	0.99495\\
0.0059	0.9951\\
0.00505	0.99375\\
0.0048	0.9937\\
0.0042	0.99265\\
0.00365	0.99225\\
0.00325	0.99175\\
0.0029	0.99195\\
0.0028	0.9899\\
0.00265	0.991\\
0.0021	0.99035\\
0.00195	0.9887\\
0.0017	0.9878\\
0.0015	0.98885\\
0.00135	0.98605\\
0.0012	0.98565\\
0.00115	0.98605\\
0.0012	0.9848\\
0.00115	0.98475\\
0.00105	0.98405\\
0.001	0.9842\\
0.0009	0.9848\\
0.00085	0.9833\\
0.0008	0.98215\\
0.0007	0.98345\\
0.0006	0.9817\\
0.0006	0.9815\\
0.00055	0.9803\\
0.00055	0.97905\\
0.0005	0.97885\\
0.0004	0.9781\\
0.00035	0.97655\\
0.00025	0.9766\\
0.00025	0.97475\\
0.00025	0.97265\\
0.00025	0.97155\\
0.0002	0.97035\\
0.0002	0.96485\\
0.0002	0.96155\\
0.0002	0.95905\\
0.0002	0.9555\\
0.0002	0.95135\\
0.0002	0.9479\\
0.00015	0.9443\\
0.00015	0.93845\\
0.00015	0.93165\\
0.00015	0.92465\\
0.00015	0.9179\\
0.0001	0.9087\\
0.0001	0.89855\\
0.0001	0.8876\\
5e-05	0.87495\\
5e-05	0.85875\\
5e-05	0.842\\
5e-05	0.8227\\
5e-05	0.80165\\
5e-05	0.77925\\
5e-05	0.75345\\
5e-05	0.7242\\
5e-05	0.6924\\
5e-05	0.6601\\
5e-05	0.61965\\
5e-05	0.57415\\
5e-05	0.52335\\
5e-05	0.46545\\
5e-05	0.402\\
5e-05	0.33865\\
5e-05	0.27385\\
0	0.2081\\
0	0.1452\\
0	0.09505\\
0	0.0562\\
0	0.03135\\
0	0.0159\\
0	0.00635\\
0	0.0034\\
0	0.00145\\
0	0.0005\\
0	0.0005\\
0	0.00045\\
0	0.00055\\
0	0.0007\\
0	0.0006\\
0	0.0006\\
0	0.00055\\
0	0.00045\\
0	0.00085\\
0	0.0004\\
0	0.00055\\
0	0.0005\\
0	0.0008\\
0	0.0005\\
0	0.00065\\
0	0.0003\\
0	0.00065\\
0	0.00095\\
0	0.0006\\
0	0.00075\\
0	0.0009\\
0	0.0007\\
0	0.0004\\
0	0.00115\\
0	0.00105\\
0	0.00045\\
0	0.00065\\
0	0.00085\\
0	0.0011\\
0	0.0006\\
0	0.00025\\
0	0.0006\\
0	0.00055\\
0	0.00035\\
0	0.00065\\
0	0.00075\\
0	0.0009\\
0	0.00045\\
0	0.00055\\
0	0.0007\\
0	0.0006\\
0	0.0007\\
0	0.0006\\
0	0.0007\\
0	0.00055\\
0	0.0008\\
0	0.0006\\
0	0.0005\\
0	0.00035\\
0	0.0005\\
0	0.00065\\
0	0.0004\\
0	0.0007\\
0	0.0008\\
0	0.0006\\
0	0.0008\\
0	0.0004\\
0	0.00055\\
0	0.00075\\
0	0.00065\\
0	0.00045\\
0	0.0006\\
0	0.0008\\
0	0.00045\\
0	0.00055\\
0	0.0005\\
0	0.00075\\
0	0.0007\\
0	0.00045\\
0	0.0004\\
0	0.00045\\
0	0.0007\\
0	0.00035\\
0	0.0008\\
0	0.0008\\
0	0.0004\\
0	0.00115\\
0	0.0008\\
0	0.00055\\
0	0.0004\\
0	0.0006\\
0	0.0007\\
0	0.00085\\
0	0.0006\\
0	0.0004\\
0	0.00075\\
0	0.00065\\
0	0.00035\\
0	0.00055\\
0	0.00065\\
0	0.0005\\
0	0.0004\\
0	0.0006\\
0	0.00055\\
0	0.0005\\
0	0.00075\\
0	0.0003\\
0	0.0005\\
0	0.00035\\
0	0.0004\\
0	0.00065\\
0	0.00055\\
0	0.0009\\
0	0.0009\\
0	0.00095\\
0	0.0006\\
0	0.00045\\
0	0.00045\\
0	0.0005\\
0	0.00055\\
0	0.00065\\
0	0.00035\\
0	0.00045\\
0	0.00035\\
0	0.00035\\
0	0.00075\\
0	0.0006\\
0	0.0004\\
0	0.0004\\
0	0.00075\\
0	0.00055\\
0	0.0005\\
0	0.0008\\
0	0.00045\\
0	0.0004\\
0	0.00045\\
0	0.00045\\
0	0.0006\\
0	0.00055\\
0	0.00085\\
0	0.00065\\
0	0.00035\\
0	0.00035\\
0	0.00055\\
0	0.0006\\
0	0.0004\\
0	0.00085\\
0	0.0006\\
0	0.00045\\
0	0.0005\\
0	0.0004\\
0	0.0003\\
0	0.00075\\
0	0.00085\\
0	0.0008\\
0	0.0006\\
0	0.0006\\
0	0.00035\\
0	0.0007\\
0	0.00075\\
0	0.0003\\
0	0.00065\\
0	0.0006\\
0	0.00065\\
0	0.0004\\
0	0.0012\\
0	0.00075\\
0	0.0006\\
0	0.0004\\
0	0.0006\\
0	0.00045\\
0	0.00095\\
0	0.0007\\
0	0.0006\\
0	0.0009\\
0	0.00055\\
0	0.0005\\
0	0.00065\\
0	0.00075\\
0	0.0004\\
0	0.0006\\
0	0.0004\\
0	0.0006\\
0	0.0003\\
0	0.0004\\
0	0.00065\\
0	0.00035\\
0	0.0005\\
0	0.00075\\
0	0.00065\\
0	0.0007\\
0	0.00065\\
0	0.0005\\
0	0.0006\\
0	0.001\\
0	0.0008\\
0	0.0004\\
0	0.0007\\
0	0.00055\\
0	0.00065\\
0	0.00085\\
0	0.00055\\
0	0.0007\\
0	0.00065\\
0	0.0004\\
0	0.0006\\
0	0.0004\\
0	0.00025\\
0	0.00045\\
0	0.00055\\
0	0.0006\\
0	0.00045\\
0	0.00065\\
0	0.00045\\
0	0.0005\\
0	0.0006\\
0	0.0005\\
0	0.0004\\
0	0.0006\\
0	0.00055\\
0	0.00065\\
0	0.0005\\
0	0.0008\\
0	0.00075\\
0	0.0007\\
0	0.0007\\
0	0.0005\\
0	0.00085\\
0	0.0006\\
0	0.00085\\
0	0.0007\\
0	0.0005\\
0	0.00065\\
0	0.00035\\
0	0.0006\\
0	0.0008\\
0	0.00055\\
0	0.0005\\
0	0.0004\\
0	0.00075\\
0	0.0003\\
0	0.0003\\
0	0.00065\\
0	0.00055\\
0	0.0004\\
0	0.0008\\
0	0.0007\\
0	0.0004\\
0	0.0004\\
0	0.0008\\
0	0.0004\\
0	0.00045\\
0	0.0004\\
0	0.00045\\
0	0.00045\\
0	0.0005\\
0	0.00055\\
0	0.0004\\
0	0.00055\\
0	0.00015\\
0	0.0007\\
0	0.00035\\
0	0.0007\\
0	0.00055\\
0	0.0006\\
0	0.0007\\
0	0.00065\\
0	0.00065\\
0	0.00065\\
0	0.00065\\
0	0.00065\\
0	0.00065\\
0	0.00045\\
0	0.0003\\
0	0.00025\\
0	0.00045\\
0	0.0002\\
0	0.00025\\
0	0.0009\\
0	0.00055\\
0	0.00055\\
0	0.0007\\
0	0.0008\\
0	0.0007\\
0	0.0004\\
0	0.00075\\
0	0.0008\\
0	0.00065\\
0	0.0004\\
0	0.00085\\
0	0.00065\\
0	0.00055\\
0	0.0005\\
0	0.0006\\
0	0.00055\\
0	0.0008\\
0	0.00055\\
0	0.00055\\
0	0.00045\\
0	0.00055\\
0	0.0005\\
0	0.00065\\
0	0.0007\\
0	0.00065\\
0	0.0004\\
0	0.0003\\
0	0.00045\\
0	0.0006\\
0	0.00065\\
0	0.00075\\
0	0.00045\\
0	0.0004\\
0	0.00065\\
0	0.00045\\
0	0.0005\\
0	0.0005\\
0	0.0005\\
0	0.00055\\
0	0.00045\\
0	0.0007\\
0	0.00045\\
0	0.0005\\
0	0.00085\\
0	0.00055\\
0	0.00045\\
0	0.00065\\
0	0.00075\\
0	0.0003\\
0	0.0005\\
0	0.00075\\
0	0.0005\\
0	0.0005\\
0	0.00035\\
0	0.00045\\
0	0.00085\\
0	0.00065\\
0	0.00055\\
0	0.0003\\
0	0.00055\\
0	0.0006\\
0	0.00085\\
0	0.00055\\
0	0.0005\\
0	0.0005\\
0	0.00055\\
0	0.0006\\
0	0.0003\\
0	0.00045\\
0	0.00085\\
0	0.00065\\
0	0.0005\\
0	0.0009\\
0	0.00065\\
0	0.00055\\
0	0.00065\\
0	0.00055\\
0	0.0005\\
0	0.0007\\
0	0.0006\\
0	0.00045\\
0	0.0004\\
0	0.00075\\
0	0.0007\\
0	0.0006\\
0	0.0006\\
0	0.00055\\
0	0.00065\\
0	0.00055\\
0	0.00055\\
0	0.0008\\
0	0.0005\\
0	0.00055\\
0	0.00085\\
0	0.00055\\
0	0.0007\\
0	0.00045\\
0	0.00085\\
0	0.0008\\
0	0.00085\\
0	0.0005\\
0	0.00045\\
0	0.0008\\
0	0.0004\\
0	0.00045\\
0	0.0006\\
0	0.0006\\
0	0.0005\\
0	0.00065\\
0	0.00075\\
0	0.0004\\
0	0.00055\\
0	0.0007\\
0	0.00055\\
0	0.00055\\
0	0.0004\\
0	0.0005\\
0	0.00025\\
0	0.00045\\
0	0.00095\\
0	0.0004\\
0	0.00025\\
0	0.0007\\
0	0.00075\\
0	0.0006\\
0	0.0005\\
0	0.0005\\
0	0.0007\\
0	0.00035\\
0	0.00045\\
0	0.00045\\
0	0.00065\\
0	0.00065\\
0	0.0006\\
0	0.0008\\
0	0.0005\\
0	0.00055\\
0	0.0004\\
0	0.00035\\
0	0.0006\\
0	0.00045\\
0	0.0007\\
0	0.00065\\
0	0.00055\\
0	0.0005\\
0	0.0005\\
0	0.00055\\
0	0.0007\\
0	0.00065\\
0	0.00055\\
0	0.00055\\
0	0.00055\\
0	0.0006\\
0	0.0006\\
0	0.0004\\
0	0.00045\\
0	0.00075\\
0	0.00075\\
0	0.00065\\
0	0.00035\\
0	0.0003\\
0	0.00065\\
0	0.0006\\
0	0.00025\\
0	0.00045\\
0	0.0007\\
0	0.00055\\
0	0.0005\\
0	0.00055\\
0	0.0005\\
0	0.0004\\
0	0.0006\\
0	0.00045\\
0	0.0006\\
0	0.00055\\
0	0.00055\\
0	0.00055\\
0	0.00055\\
0	0.0009\\
0	0.00075\\
0	0.00085\\
0	0.0004\\
0	0.0007\\
};

\addplot[only marks,mark=x,mark size=4pt,cyan]
  table[row sep=crcr]{%
0	0.0005\\
};

\end{axis}
\end{tikzpicture}\label{fig:sis_phase1}}
      \subfloat[$r=3 z$]{\input{sis2_phase}\label{fig:sis_phase2}}
      \subfloat[$r=3 z^2$]{\input{sis3_phase}\label{fig:sis_phase3}}
      \subfloat[$r=3 z$, no social influence]{\input{sis4_phase}\label{fig:sis_phase4}\qquad\quad}
    \caption{Simulations of the SIS model. In (a)--(d), we show the time-evolution of the  epidemic prevalence $z(t)$ (red) and the fraction of adopters of self-protections $\langle x(t)\rangle:=\frac1n\sum_{i\in\mathcal V}x_i(t)$ (intensity of the blue bands). In (e)--(h), we show the corresponding trajectories on the phase-space. Panels refer to different risk perceptions $r(z)$, as detailed in the sub-captions.    }
    \label{fig:sis}
\end{figure*}

In the following, we opt for modeling the contact layer $\mathcal E_C(t)$ by means of a discrete-time activity-driven network (ADN)~\cite{Perra2012}, which has found successful application in mathematical epidemiology~\cite{Rizzo2016Ebola}. In ADNs, each individual $i\in\mathcal V$ is characterized by a constant parameter  $a_i\in[0,1]$, called \emph{activity}, which quantifies their probability to be ``active'' and thus generate a fixed number $m \geq 1$ of undirected links with other individuals selected uniformly at random from the population, for each discrete-time step $t$. Similar to its original formulation, we choose $m$ to be constant and equal for all individuals. These contacts are added to the link set $\mathcal E_C(t)$, contribute to the epidemic process, and are then removed before the next discrete time instant and the next activation of individuals. Despite their simplicity, which enables rigorous analytical treatment and fast numerical simulations~\cite{Perra2012,Zino2016}, ADNs can capture important features of complexity that characterize real-world networks, including their temporal and heterogeneous nature, and further features can be incorporated in an analytically-tractable manner~\cite{Pozzana2017,zino2018memory,Petri2018,Leitch2019}.

\subsection*{Epidemic threshold}

For large populations and fully connected influence layers, we compute the epidemic threshold via a mean-field approach~\cite{VanMieghem2009}. In the absence of cumulative frustration ($\gamma=0$), which is a reasonable assumption in the early stages of an outbreak, and assuming constant policy interventions $u(t)=\bar u$, the outbreak is quickly eradicated if
    \begin{equation}\label{eq:threshold}
\frac{\lambda}{\mu}<\frac{e^{\beta(1-\bar u)}+(1-\beta) e^{-\beta c}}{m(\langle a\rangle+\sqrt{\langle a^2\rangle})(e^{\beta(1-\bar u)}+(1-\beta-\sigma) e^{-\beta c})}\,,
\end{equation}
where $\langle a\rangle:=\frac1n\sum_{i\in\mathcal V}a_i$ and $\langle a^2\rangle:=\frac1n\sum_{i\in\mathcal V}a^2_i$ are the mean and second moment of the activity distribution, respectively (the derivation of~\eqref{eq:threshold} is reported in Appendix~\ref{app:a}). Note that, when the cost for adopting self-protections grows large, $c\to\infty$, the threshold in \eqref{eq:threshold} tends to the same threshold expression as that of a standard SIS model on ADNs, which is $\lambda/\mu<(m(\langle a\rangle+\sqrt{\langle a^2\rangle}))^{-1}$~\cite{Perra2012}. 

The threshold in~\eqref{eq:threshold} offers insight into the role of human behavior in the early stages of an epidemic outbreak by establishing  conditions under which the disease is immediately eradicated. However, the key novelty of the proposed paradigm lies in the possibility to investigate the interplay between  human behavior and epidemic spreading in the long term, when the epidemic actually spreads (i.e., above the epidemic threshold). In this scenario, such a complex interplay may give rise to several interesting and realistic phenomena, such as periodic oscillations and multiple waves, the emergence of endemic diseases, and even behavioral responses leading to the successful eradication of the outbreak. In the following, we execute a numerical study to elucidate such phenomena~\footnote{The code is available at {https://github.com/lzino90/behavior}.}. Specifically, we combine our co-evolutionary model with the classical SIS and susceptible--infected--recovered (SIR) models, parametrized to reproduce real-world diseases. The SIS model is used to illustrate the large variety of phenomena that the paradigm can reproduce, and to discuss the key role of social influence into determining a collective response, confirming recent empirical evidence from the social psychology literature~\cite{Tuncgenc2021influence}. The SIR model is utilized to show that our model can reproduce real-world epidemic patterns, and to discuss the design of policy interventions. In all the simulations, we fix $n = 20,000$ individuals ($20$ of them initially infected, selected uniformly at random). The influence layer is modeled through a Watts--Strogatz small-world network~\cite{Watts1998} with average degree $8$ and rewiring probability $1/8$. The contact layer is generated through an ADN with power-law distributed activities $a_i$ with a negative exponent $-2.09$, as in~\cite{Aiello2001}, and lower cutoff at $a_{\min}=0.1$.

\subsection*{SIS model}

We simulate an SIS model calibrated on gonorrhea~\cite{Yorke1978} (see Appendix~\ref{app:b} and Table~\ref{tab:parameters}), by fixing an immediate cost of $c=0.3$ with no accumulation $\gamma=0$ and assuming that no policy interventions are enacted ($u(t)=0,\,\forall t\geq 0$). In Figs.~\ref{fig:sis}a--\ref{fig:sis}c, we consider three scenarios with progressively less cautious populations ($r(z)=3\sqrt{z}$, $r(z)=3{z}$, and $r(z)=3z^2$, respectively, see Appendix~\ref{app:b}). We observe that this shift in risk perception causes not only a quantitative shift in the epidemic dynamics, similar to most awareness-based models~\cite{Sahneh2012,Granell2013}, but more importantly, qualitatively changes the salient phenomena. In fact, the phenomena spans from a prompt and sustained collective response that leads to fast eradication of the disease (in Fig.~\ref{fig:sis_phase1}, the trajectory rapidly reaches the disease-free manifold $z=0$), to periodic oscillations both in the epidemic prevalence and in the behavioral response (Figs.~\ref{fig:sis2} and~\ref{fig:sis_phase2}), and finally to a partial behavioral response, resulting in the emergence of a meta-stable endemic equilibrium (Figs.~\ref{fig:sis3} and~\ref{fig:sis_phase3}). The periodic oscillations are similar to that observed in several existing works, such as~\cite{Just2017,Steinegger2020}.
 
In the scenario depicted above, the risk perception function determines a critical prevalence $z^*=\min\{z:r(z)>1+c\}$, such that $z(t)>z^*$ implies that the payoff for adopting self-protections is larger than for not adopting them ($\pi_i^1(t)>\pi_i^0(t)$), for any individual, irrespective of the behavior of others. With $r(z)=3\sqrt z$, the critical prevalence is $z^*\approx18\%$. However, as can be observed in Fig.~\ref{fig:sis_phase1}, social influence causes individuals to rapidly and widely adopt self-protective behaviors at a much earlier prevalence of $z\approx 6\%$, highlighting the key role played by social influence toward facilitating the emergence of collective behavioral patterns and, in this instance, helping in the fast eradication of the disease. 

We further investigate the role of social influence by simulating the model in its absence (i.e., removing the first term in Eqs.~(\ref{eq:pi0})--(\ref{eq:pi1}), as detailed in Appendix C). Our findings support the intuition that social influence is key to ensure collective population responses, which are in turn crucial for the successful eradication of the disease (Fig.~\ref{fig:sis1}) and the emergence of periodic oscillations (Fig.~\ref{fig:sis2}). In fact, in the absence of social influence, the system shows a less rich range of behaviors, whereby a partial behavioral response always leads to the convergence to an endemic equilibrium (Figs.~\ref{fig:sis4} and~\ref{fig:sis_phase4} are obtained with the parameters of Fig.~\ref{fig:sis2} without social influence. However, interestingly, social influence can also cause collective rejection of self-protective behaviors, and this results in a higher peak disease prevalence with respect to the scenario without social influence. Simulations corresponding to Figs.~\ref{fig:sis1} and~\ref{fig:sis3} are reported in Appendix C (Fig.~\ref{fig:sis_noSI}).

\subsection*{SIR model}

\begin{figure*}
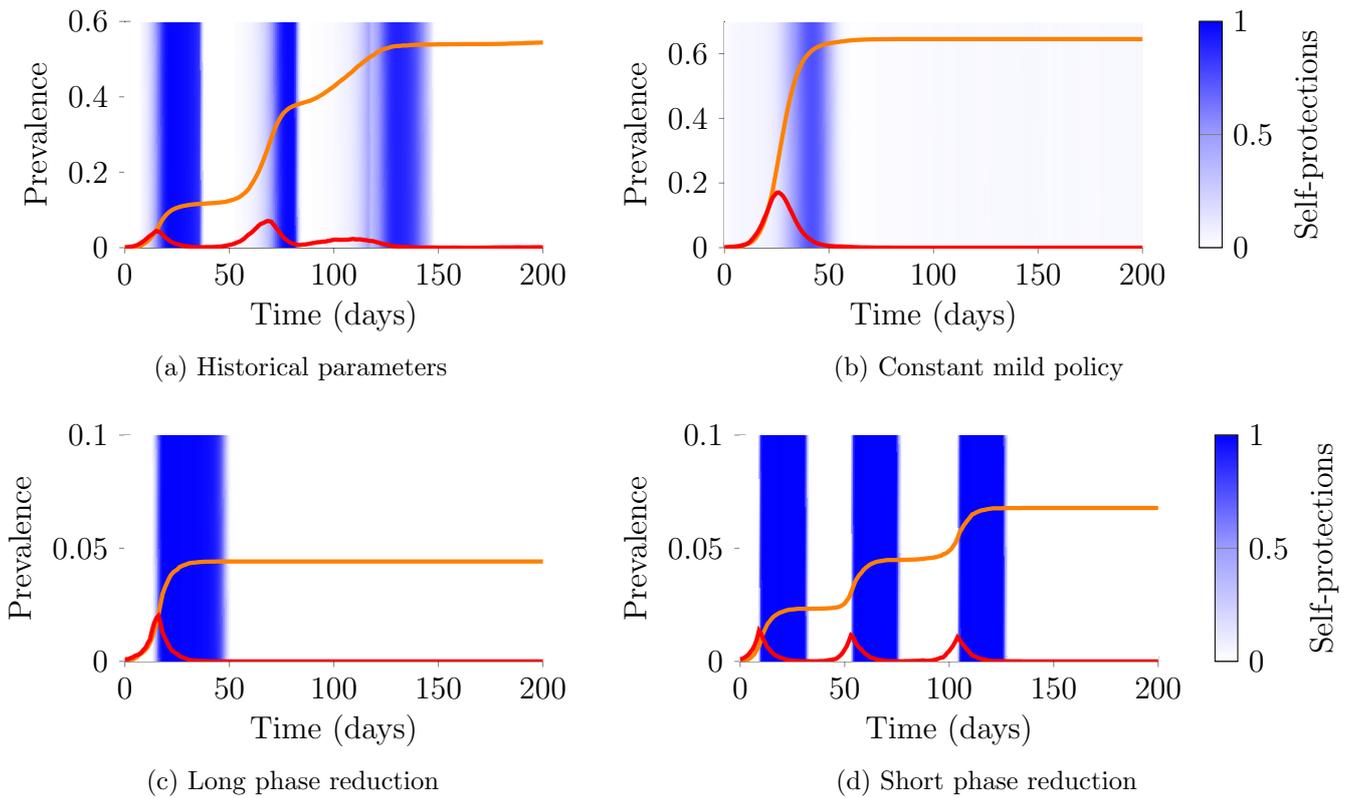

    \centering
        \subfloat[Historical parameters]{\input{flu1}\label{fig:flu1}}\,\,\,
    \subfloat[Constant mild policy]{\input{flu2}\label{fig:flu2}}\\
    \subfloat[Long phase reduction]{\begin{tikzpicture}
\begin{axis}[%
axis lines=left,
x    axis line style={-},
y    axis line style={-},
width=\lf cm,
height=\hf cm,
point meta min=0, 
   point meta max=1,
scale only axis,
xmin=0,
xmax=200,
ymin=0,
ytick distance=.05,
xlabel shift={-2pt},
ymax=.1,
ylabel={Prevalence},
xlabel={Time (days)},
]

 \addplot [on layer=axis background,
                surf,
                point meta=explicit,
                shader=interp,
               draw=none,     
            ]
            table [x=x,y=y,meta=z] {
                x   y   z
0	0 0
1	0 0.0025
2	0 0.0024
3	0 0.00195
4	0 0.00205
5	0 0.0019
6	0 0.00215
7	0 0.0024
8	0 0.00155
9	0 0.00265
10	0 0.00205
11	0 0.0017
12	0 0.00225
13	0 0.0023
14	0 0.10835
15	0 0.2948
16	0 0.6031
17	0 0.8892
18	0 0.98985
19	0 0.99945
20	0 0.9992
21	0 0.9992
22	0 0.9988
23	0 0.99865
24	0 0.9978
25	0 0.9975
26	0 0.9972
27	0 0.9967
28	0 0.9953
29	0 0.9944
30	0 0.99325
31	0 0.9921
32	0 0.9909
33	0 0.989
34	0 0.9887
35	0 0.9872
36	0 0.98555
37	0 0.98235
38	0 0.97755
39	0 0.97205
40	0 0.96615
41	0 0.9566
42	0 0.9416
43	0 0.9195
44	0 0.8856
45	0 0.83455
46	0 0.7583
47	0 0.63745
48	0 0.46445
49	0 0.2643
50	0 0.0998
51	0 0.02
52	0 0.00255
53	0 0.00155
54	0 0.00125
55	0 0.0017
56	0 0.00125
57	0 0.00215
58	0 0.0024
59	0 0.0023
60	0 0.0028
61	0 0.00235
62	0 0.0033
63	0 0.00345
64	0 0.00295
65	0 0.00295
66	0 0.00245
67	0 0.0024
68	0 0.00225
69	0 0.00215
70	0 0.002
71	0 0.00235
72	0 0.00235
73	0 0.0016
74	0 0.00225
75	0 0.0011
76	0 0.00155
77	0 0.00145
78	0 0.0011
79	0 0.00155
80	0 0.0019
81	0 0.00165
82	0 0.0014
83	0 0.00175
84	0 0.0018
85	0 0.00195
86	0 0.002
87	0 0.0023
88	0 0.0017
89	0 0.0015
90	0 0.0011
91	0 0.00125
92	0 0.0019
93	0 0.00205
94	0 0.0017
95	0 0.00165
96	0 0.002
97	0 0.00175
98	0 0.0019
99	0 0.0018
100	0 0.0017
101	0 0.00115
102	0 0.0015
103	0 0.00175
104	0 0.00205
105	0 0.00195
106	0 0.00145
107	0 0.0018
108	0 0.002
109	0 0.00155
110	0 0.002
111	0 0.0015
112	0 0.00125
113	0 0.00125
114	0 0.00175
115	0 0.0018
116	0 0.00225
117	0 0.00165
118	0 0.00225
119	0 0.00205
120	0 0.0022
121	0 0.0023
122	0 0.00195
123	0 0.00135
124	0 0.0023
125	0 0.002
126	0 0.00175
127	0 0.00235
128	0 0.0017
129	0 0.00145
130	0 0.00175
131	0 0.00155
132	0 0.00165
133	0 0.0021
134	0 0.00195
135	0 0.0021
136	0 0.00165
137	0 0.00155
138	0 0.00175
139	0 0.0018
140	0 0.00205
141	0 0.0018
142	0 0.00175
143	0 0.0022
144	0 0.0019
145	0 0.00225
146	0 0.00225
147	0 0.00165
148	0 0.00205
149	0 0.0022
150	0 0.00175
151	0 0.00215
152	0 0.00195
153	0 0.00205
154	0 0.00175
155	0 0.00215
156	0 0.0022
157	0 0.00195
158	0 0.00165
159	0 0.00165
160	0 0.00185
161	0 0.0014
162	0 0.0019
163	0 0.00215
164	0 0.00125
165	0 0.0015
166	0 0.00125
167	0 0.002
168	0 0.0015
169	0 0.0013
170	0 0.0022
171	0 0.00165
172	0 0.0025
173	0 0.0026
174	0 0.00245
175	0 0.00215
176	0 0.00175
177	0 0.00185
178	0 0.00195
179	0 0.0015
180	0 0.00165
181	0 0.00165
182	0 0.00165
183	0 0.00225
184	0 0.0024
185	0 0.00175
186	0 0.0013
187	0 0.0015
188	0 0.0019
189	0 0.00175
190	0 0.00225
191	0 0.00195
192	0 0.00275
193	0 0.00205
194	0 0.00155
195	0 0.0016
196	0 0.00205
197	0 0.0019
198	0 0.00195
199	0 0.0021
200	0 0.00205

0	0.1 0
1	0.1 0.0025
2	0.1 0.0024
3	0.1 0.00195
4	0.1 0.00205
5	0.1 0.0019
6	0.1 0.00215
7	0.1 0.0024
8	0.1 0.00155
9	0.1 0.00265
10	0.1 0.00205
11	0.1 0.0017
12	0.1 0.00225
13	0.1 0.0023
14	0.1 0.10835
15	0.1 0.2948
16	0.1 0.6031
17	0.1 0.8892
18	0.1 0.98985
19	0.1 0.99945
20	0.1 0.9992
21	0.1 0.9992
22	0.1 0.9988
23	0.1 0.99865
24	0.1 0.9978
25	0.1 0.9975
26	0.1 0.9972
27	0.1 0.9967
28	0.1 0.9953
29	0.1 0.9944
30	0.1 0.99325
31	0.1 0.9921
32	0.1 0.9909
33	0.1 0.989
34	0.1 0.9887
35	0.1 0.9872
36	0.1 0.98555
37	0.1 0.98235
38	0.1 0.97755
39	0.1 0.97205
40	0.1 0.96615
41	0.1 0.9566
42	0.1 0.9416
43	0.1 0.9195
44	0.1 0.8856
45	0.1 0.83455
46	0.1 0.7583
47	0.1 0.63745
48	0.1 0.46445
49	0.1 0.2643
50	0.1 0.0998
51	0.1 0.02
52	0.1 0.00255
53	0.1 0.00155
54	0.1 0.00125
55	0.1 0.0017
56	0.1 0.00125
57	0.1 0.00215
58	0.1 0.0024
59	0.1 0.0023
60	0.1 0.0028
61	0.1 0.00235
62	0.1 0.0033
63	0.1 0.00345
64	0.1 0.00295
65	0.1 0.00295
66	0.1 0.00245
67	0.1 0.0024
68	0.1 0.00225
69	0.1 0.00215
70	0.1 0.002
71	0.1 0.00235
72	0.1 0.00235
73	0.1 0.0016
74	0.1 0.00225
75	0.1 0.0011
76	0.1 0.00155
77	0.1 0.00145
78	0.1 0.0011
79	0.1 0.00155
80	0.1 0.0019
81	0.1 0.00165
82	0.1 0.0014
83	0.1 0.00175
84	0.1 0.0018
85	0.1 0.00195
86	0.1 0.002
87	0.1 0.0023
88	0.1 0.0017
89	0.1 0.0015
90	0.1 0.0011
91	0.1 0.00125
92	0.1 0.0019
93	0.1 0.00205
94	0.1 0.0017
95	0.1 0.00165
96	0.1 0.002
97	0.1 0.00175
98	0.1 0.0019
99	0.1 0.0018
100	0.1 0.0017
101	0.1 0.00115
102	0.1 0.0015
103	0.1 0.00175
104	0.1 0.00205
105	0.1 0.00195
106	0.1 0.00145
107	0.1 0.0018
108	0.1 0.002
109	0.1 0.00155
110	0.1 0.002
111	0.1 0.0015
112	0.1 0.00125
113	0.1 0.00125
114	0.1 0.00175
115	0.1 0.0018
116	0.1 0.00225
117	0.1 0.00165
118	0.1 0.00225
119	0.1 0.00205
120	0.1 0.0022
121	0.1 0.0023
122	0.1 0.00195
123	0.1 0.00135
124	0.1 0.0023
125	0.1 0.002
126	0.1 0.00175
127	0.1 0.00235
128	0.1 0.0017
129	0.1 0.00145
130	0.1 0.00175
131	0.1 0.00155
132	0.1 0.00165
133	0.1 0.0021
134	0.1 0.00195
135	0.1 0.0021
136	0.1 0.00165
137	0.1 0.00155
138	0.1 0.00175
139	0.1 0.0018
140	0.1 0.00205
141	0.1 0.0018
142	0.1 0.00175
143	0.1 0.0022
144	0.1 0.0019
145	0.1 0.00225
146	0.1 0.00225
147	0.1 0.00165
148	0.1 0.00205
149	0.1 0.0022
150	0.1 0.00175
151	0.1 0.00215
152	0.1 0.00195
153	0.1 0.00205
154	0.1 0.00175
155	0.1 0.00215
156	0.1 0.0022
157	0.1 0.00195
158	0.1 0.00165
159	0.1 0.00165
160	0.1 0.00185
161	0.1 0.0014
162	0.1 0.0019
163	0.1 0.00215
164	0.1 0.00125
165	0.1 0.0015
166	0.1 0.00125
167	0.1 0.002
168	0.1 0.0015
169	0.1 0.0013
170	0.1 0.0022
171	0.1 0.00165
172	0.1 0.0025
173	0.1 0.0026
174	0.1 0.00245
175	0.1 0.00215
176	0.1 0.00175
177	0.1 0.00185
178	0.1 0.00195
179	0.1 0.0015
180	0.1 0.00165
181	0.1 0.00165
182	0.1 0.00165
183	0.1 0.00225
184	0.1 0.0024
185	0.1 0.00175
186	0.1 0.0013
187	0.1 0.0015
188	0.1 0.0019
189	0.1 0.00175
190	0.1 0.00225
191	0.1 0.00195
192	0.1 0.00275
193	0.1 0.00205
194	0.1 0.00155
195	0.1 0.0016
196	0.1 0.00205
197	0.1 0.0019
198	0.1 0.00195
199	0.1 0.0021
200	0.1 0.00205
            };
        
\addplot [color=orange,ultra thick]
  table[row sep=crcr]{%
0	0\\
1	0.00015\\
2	0.00035\\
3	0.00055\\
4	0.00095\\
5	0.0014\\
6	0.0021\\
7	0.00295\\
8	0.00355\\
9	0.0043\\
10	0.00525\\
11	0.0064\\
12	0.0083\\
13	0.0103\\
14	0.0132\\
15	0.01695\\
16	0.02065\\
17	0.0251\\
18	0.0296\\
19	0.03235\\
20	0.0344\\
21	0.03645\\
22	0.03815\\
23	0.03925\\
24	0.04\\
25	0.04055\\
26	0.04155\\
27	0.04205\\
28	0.04245\\
29	0.04275\\
30	0.0431\\
31	0.04345\\
32	0.04355\\
33	0.04365\\
34	0.0437\\
35	0.04385\\
36	0.04395\\
37	0.04395\\
38	0.04405\\
39	0.04405\\
40	0.04405\\
41	0.0441\\
42	0.0441\\
43	0.0441\\
44	0.04415\\
45	0.04415\\
46	0.04415\\
47	0.04415\\
48	0.04415\\
49	0.04415\\
50	0.04415\\
51	0.04415\\
52	0.04415\\
53	0.04415\\
54	0.04415\\
55	0.04415\\
56	0.04415\\
57	0.04415\\
58	0.04415\\
59	0.04415\\
60	0.04415\\
61	0.04415\\
62	0.04415\\
63	0.04415\\
64	0.04415\\
65	0.04415\\
66	0.04415\\
67	0.04415\\
68	0.04415\\
69	0.04415\\
70	0.04415\\
71	0.04415\\
72	0.04415\\
73	0.04415\\
74	0.04415\\
75	0.04415\\
76	0.04415\\
77	0.04415\\
78	0.04415\\
79	0.04415\\
80	0.04415\\
81	0.04415\\
82	0.04415\\
83	0.04415\\
84	0.04415\\
85	0.04415\\
86	0.04415\\
87	0.04415\\
88	0.04415\\
89	0.04415\\
90	0.04415\\
91	0.04415\\
92	0.04415\\
93	0.04415\\
94	0.04415\\
95	0.04415\\
96	0.04415\\
97	0.04415\\
98	0.04415\\
99	0.04415\\
100	0.04415\\
200	0.04415\\
};

\addplot [color=red, ultra thick]
  table[row sep=crcr]{%
0	0.001\\
1	0.00095\\
2	0.00115\\
3	0.0016\\
4	0.0019\\
5	0.00255\\
6	0.0028\\
7	0.0033\\
8	0.00405\\
9	0.00525\\
10	0.0062\\
11	0.00815\\
12	0.0097\\
13	0.0137\\
14	0.0169\\
15	0.01905\\
16	0.0201\\
17	0.017\\
18	0.01295\\
19	0.01065\\
20	0.0088\\
21	0.007\\
22	0.0054\\
23	0.0044\\
24	0.0038\\
25	0.00335\\
26	0.0024\\
27	0.002\\
28	0.0016\\
29	0.0013\\
30	0.00095\\
31	0.0006\\
32	0.0005\\
33	0.0004\\
34	0.0004\\
35	0.0003\\
36	0.0002\\
37	0.0002\\
38	0.0001\\
39	0.0001\\
40	0.0001\\
41	5e-05\\
42	5e-05\\
43	5e-05\\
44	0\\
45	0\\
46	0\\
47	0\\
48	0\\
49	0\\
50	0\\
51	0\\
52	0\\
53	0\\
54	0\\
55	0\\
56	0\\
57	0\\
58	0\\
59	0\\
60	0\\
61	0\\
62	0\\
63	0\\
64	0\\
65	0\\
66	0\\
67	0\\
68	0\\
69	0\\
70	0\\
71	0\\
72	0\\
73	0\\
74	0\\
75	0\\
76	0\\
77	0\\
78	0\\
79	0\\
80	0\\
81	0\\
82	0\\
83	0\\
84	0\\
85	0\\
86	0\\
87	0\\
88	0\\
89	0\\
90	0\\
91	0\\
92	0\\
93	0\\
94	0\\
95	0\\
96	0\\
97	0\\
98	0\\
99	0\\
100	0\\
200	0\\
	};

\end{axis}
\end{tikzpicture}\label{fig:flu3}}\,\,\,
    \subfloat[Short phase reduction]{\input{flu4}\label{fig:flu4}}
    \caption{Simulations of the SIR model. Temporal evolution of the epidemic prevalence $z(t)$ (red), recovered $R(t):=\frac1n|\{i:y_i(t)=R\}|$ (orange), and fraction of adopters of self-protections $\langle x(t)\rangle$ (intensity of the blue vertical bars) with (a) historical parameters of the 1918--19 Spanish flu pandemic~\cite{Mills2004, Markel2007}; (b) a mild and constant intervention; (c) a severe intervention with a long phased reduction  period; and (d)  a very severe intervention with a short phased reduction period. }
\end{figure*}

We further elucidate the potentialities of our behavioral paradigm by combining it with an SIR model (see Fig.~\ref{fig:network}). In the SIR model, the \emph{removed} ($R$) health state is added to $\mathcal A$ to represent immunized individuals after recovery (or death), and the system is governed by \eqref{eq:sis} and $\mathbb P[y_i(t+1)=R\,|\,y_i(t)=I]=\mu$. The model is calibrated on  the 1918--19 Spanish flu~\cite{Mills2004,Mossong2008} (see Appendix~\ref{app:b} and Table~\ref{tab:parameters}). In particular, we set a risk perception function that increases slowly, $r=3z^2$ (associated with the initial suppression of news about the flu~\cite{Barry2009}), and policy interventions are set to mimic historical lockdowns ($u(t)=0.5$ when $1\%$ of the population is infected, and held constant for $T_{a}=28$ days~\cite{Markel2007}). For airborne diseases like Spanish flu, self-protective behaviors entail physical distancing and closures of economic activities, which often yield an accumulation of psychological distress and economic losses~\cite{nicola2020socio,Qiu2020,Bartik2020economic}. Accordingly, we set $c=0.1$ and $\gamma=0.9$ (see Appendix~\ref{app:b}). Figure~\ref{fig:flu1} shows that our paradigm is able to qualitatively reproduce the historical epidemic pattern, which witnessed a resurgent pandemic with three waves that includes a massive second wave~\cite{Mills2004,Markel2007}. 

We utilize this example to discuss the design of  policy interventions. First, from Fig.~\ref{fig:flu2}, we observe that mild policy interventions, even if indefinite in duration ($u(t)=0.4,\,\forall t\geq 0$), may not be sufficient to ensure a timely and collective response, resulting in a massive outbreak that reaches more than $60\%$ of the population. Next, we consider two scenarios with severe but shorter policy interventions, followed by a linear phased reduction (Figs.~\ref{fig:flu3} and~\ref{fig:flu4}); in the first scenario, the policy is less severe ($u(t)=0.7$ vs. $u(t)=1.2$ for $21$ days), but the reduction period is longer ($42$ vs. $7$ days)~\footnote{The total control efforts for the two policies, measured as the sum of $u(t)$ over the duration of a lockdown, are the same.}. Comparing the two scenarios, we conclude that, provided that the initial policy interventions are sufficiently severe to ensure collective adoption of self-protections (thus avoiding the scenario shown in Fig.~\ref{fig:flu2}), the successful eradication of the disease depends primarily on a sufficiently long phased reduction period. This avoids the multiple waves and subsequent lockdowns that would increase both the death toll and the total social-economic cost. The latter point is consistent with recent observations on the duration of policy interventions and their gradual uplifting during the COVID-19 pandemic~\cite{Lopez2020}; however, there are far more epidemic complexities and real-world challenges to consider for COVID-19. Consistent with the SIS results, the SIR outcomes underline the key role of social influence, which may act as a double-edged sword, providing either a driving force or a retarding force for the collective adoption of self-protections, depending on whether the intervention policies are sufficiently strong.

These simulations illustrate the predictive power of our paradigm, once a proper parametrization and model have been extrapolated from empirical data. In fact, existing approaches typically estimate how infection parameters, associated with the disease dynamics, are explicitly changed due to policy interventions~\cite{Perra2021}. Generally, this is a difficult task that does not explicitly account for the complexity of human behavior. In contrast, our paradigm leaves the disease dynamics untouched, and allows policy interventions to only influence the decision-making process that determines the behavioral responses, which, in turn, shape the epidemic evolution.

\section*{Conclusion}

In summary, we proposed a novel and unified individual-level modeling paradigm that captures the co-evolution of disease spreading and collective decision-making at the same time scale. Our framework encapsulates a wide range of time-varying factors that are crucial in decision-making during epidemics, from the initial outbreak to its complete eradication, including government interventions, risk perception, bounded rationality, and social influence. The framework decouples the roles of these factors ---which collectively shape the behavioral response to an epidemic outbreak--- in an intuitive way, enabling their estimation from epidemiological~\cite{cobey2020modeling}, socio-demographic~\cite{Mossong2008}, communication~\cite{oliver2020phone}, and mobility data~\cite{google}, coupled with empirical studies on social influence~\cite{Tuncgenc2021influence}, population adherence to NPIs \cite{Haug2020,Singhe2021}, risk perception~\cite{Dryhurst2020}, socio-economic impact of NPIs and emergence of distress~\cite{Qiu2020,nicola2020socio}.

Our methodology is specifically designed to be a parsimonious paradigm, adaptable to different epidemic progression models~\cite{Pastor-Satorras2015,brauer2011mathematical} and temporal network interactions~\cite{Holme2012,Holme2015,Valdano2015,zino2018memory,Koher2019}. Its simple formulation allows to perform fast simulations, scalable to large-scale systems. Complex real-world phenomena are reproduced within our unified framework ---including periodic oscillations, multiple epidemic waves, and prompt collective response by the population leading to the fast eradication of the disease triggered by endogenous risk perception or exogenous policy interventions. In particular, the paradigm allows to investigate the impact of different policy interventions on the collective behaviors and, in turn, on mitigating the spread. 

Due to its flexibility, further features of the behavioral response can be directly incorporated in the paradigm, including mutual protection, delays in the behavioral response, and biased risk perceptions due to imperfect information on the epidemic prevalence. Finally, the combination of the proposed paradigm with more complex and realistic epidemic progression models (e.g., those tailored to COVID-19~\cite{estrada2020covid,arenas2020covid}) is key to further investigating the features captured by our model. This may allow our paradigm to be utilized to predict the behavioral response to real-world epidemic outbreaks and, thus, help public health authorities design effective interventions.

\section*{Acknowledgments}
The work by M.Y. is supported by the Western Australian Government, under the Premier's Science Fellowship Program; 
L.Z. and M.C. are partially supported by the European Research Council (ERC-CoG-771687) and the Netherlands Organization for Scientific Research (NWO-vidi-14134); A.R. by Compagnia di San Paolo and the Italian Ministry of Foreign Affairs and International Cooperation (``Mac2Mic'').

\section*{Contribution Statement}
M.Y. and L.Z. contributed equally as leading authors. A.R. and M.C. contributed equally.

\appendix

\section{Derivation of \eqref{eq:threshold}}\label{app:a}

In the absence of cumulative frustration and for a fully mixed influence layer, we observe that \eqref{eq:log-linear} has the same expression for all the individuals. In fact, if we define $\bar x(z)$ as the probability that a generic node adopts self-protective behaviors when the epidemic prevalence is equal to $z$, then, according to the strong law of large numbers, $\pi_i^0(t)=1-\bar x(z)-u(t)=1-\bar x(z)-\bar u$ (since the control is assumed to be constant) and $\pi_i^1(t)=\bar x(z)+r(z)-c$, which are independent of $i$. In a mean-field approach~\cite{VanMieghem2009}, we define $\theta(t)=\frac 1n\sum_{i:y_i(t)=I} a_i$ as the average activity of infected nodes and $z_i(t)=\mathbb P[y_i(t)=I]$. Due to the strong law of large numbers, in the limit of large-scale systems $n\to\infty$, $z(t)=\frac 1n\sum_{i=1}^n z_i(t)$ and $\theta(t)=\frac 1n\sum_{i=1}^n a_iz_i(t)$. Hence, from the mean-field evolution of $z_i(t)$, given by
\begin{equation}
    z_i(t+1)=z_i(t)-\mu z_i(t)+m\lambda (1-z_i(t))(1-\bar x(z(t))) a_i z_i(t)+(1-z_i(t))(1-\bar x(z(t)))\lambda m \theta(t),
\end{equation}
we determine the following system of difference equations for the epidemic prevalence and the average activity of infected individuals, linearized about the disease-free equilibrium ($z=0$, $\theta=0$):
\begin{equation}\label{eq:dynamics}
\begin{array}{lll}
 z(t+1)&=&z(t)-\mu z(t)+m\lambda \langle a\rangle z(t)(1-\sigma \bar x(0))+m\lambda (1-\sigma \bar x(0))\theta(t)\\[5pt]
 \theta(t+1)&=&\theta(t)-\mu \theta(t)+m\lambda \langle a^2\rangle z(t)(1-\sigma\bar x(0))+m\lambda  \langle a\rangle(1-\sigma \bar x(0))\theta(t)\,,
\end{array}
\end{equation}
where $\langle a\rangle$ and $\langle a^2\rangle$ are the average and second moment of the activity distribution, respectively.

From standard theory on the stability of discrete-time linear time-invariant systems~\cite{rugh1996linear}, the origin is stable if
\begin{equation}\label{eq:threshold-1}
\frac{\lambda}{\mu}<\frac{1}{m(\langle a\rangle+\sqrt{\langle a^2\rangle})(1-\sigma\bar x(0))}\,.
\end{equation}

In fully-mixed influence layers, the probability for an individual to adopt self-protective behaviors $\bar x(z)$ can be derived by substituting $\pi_i(0)=1-\bar x(z)-\bar u$ and $\pi_i(1)=\bar x(z)+r(z)-c$ into~\eqref{eq:log-linear}, obtaining the equilibrium equation:
\begin{equation}
\bar x=\frac{e^{\beta (\bar x-c+r(z))}}{e^{\beta (\bar x-c+r(z))}+e^{\beta (1-\bar x-\bar u)}}\,.
\end{equation}
Even though it is not possible to derive a closed-form solution, we observe that at the inception of the epidemic outbreak, $x_i(0)=0$ for all individuals and, for sufficiently small values of $\bar u$ (i.e., $\bar u \ll 1+c$), in the early stages it is verified that $\pi_i^0(t)>\pi_i^1(t)$. Hence, if the rationality $\beta$ is sufficiently large, the equilibrium $\bar x$ is close to $0$ and can be approximated by Taylor-expanding the right-hand side of the equation, obtaining
\begin{equation}\bar x(z)\approx\frac{e^{\beta(-c+r(z))}}{e^{\beta(1-\bar u)}+(1-\beta)e^{\beta(-c+r(z))}}\,,\end{equation}
which can be evaluated for $z=0$ and inserted into \eqref{eq:threshold-1}, obtaining \eqref{eq:threshold}.

\section{Parameters used in the simulations}\label{app:b}

\subsection*{Epidemic parameters}

The SIS is calibrated on gonorrhea, which is a sexually transmitted disease characterized by negligible protective immunity after recovery and negligible latency period (individuals are infectious on average the day after contagion)~\cite{Lajmanovich1976,Yorke1978}. The SIR model is parametrized based on Spanish flu pandemic, which is characterized by a short latency period (individuals are infectious on average after 1--2 days from the contagion) that can be neglected and by protective immunity gained after recovery~\cite{Mills2004}.  

The epidemic parameters are set from epidemiological data. Specifically, reliable estimations of the time from infection to recovery $\tau$ are available~\cite{Yorke1978,Mills2004} (namely, $\tau=55$ days for gonorrhea and $\tau=4.1$ days for Spanish flu). Similar to Prem et al.~\cite{Prem2020}, from these data we define $\mu=1-\exp\left(-{1}/{\tau}\right)$. 

The parameter $\lambda$ is obtained from available estimations of the basic reproduction number $R_0$ for the two diseases~\cite{Yorke1978,Mills2004} (namely, $R_0=1.6$ for gonorrhea and $R_0=2$ for Spanish flu). The basic reproduction number is defined as the average number of secondary infections produced by an infected individual in a population where everyone is susceptible. Hence, given that  $\tau$ is the average time that an individual is infectious, assuming independence between the time an individual is infectious and their activity, we compute
\begin{equation}
    R_0=\frac{1}{n}\sum_{i\in\mathcal V}(a_i+\langle a\rangle) m \lambda \tau=2\langle a\rangle m \lambda \tau\,,
\end{equation}
which implies $\lambda={R_0}/{2m\langle a\rangle\tau}$. For the effectiveness of self-protective behaviors, we have assumed that they prevent $99\%$ of the contagions in the SIS (gonorrhea) scenario, where self-protective behavior may include the use of physical protection barriers such as condoms). For the SIR model (Spanish flu), we assume a $95\%$ effectiveness at preventing contagion, with self-protective behavior including physical distancing, stay-at-home actions, and wearing masks. 

The epidemic parameters computed using this procedure are gathered in Table~\ref{tab:parameters}. 

\begin{table}
\centering
\caption{Parameters used in the simulations.}\label{tab:parameters}
\begin{tabular}{c| c |c}
\,\,Parameter\,\,&\,\,SIS (gonorrhea)~\cite{Yorke1978} \,\,&\,\,SIR (Spanish flu)~\cite{Mills2004}\,\,\\
 \hline
$\lambda$&0.3626&0.066\\
$\mu$&0.1195&0.2164\\
$\sigma$&0.99&0.95\\
time unit&week&day
\end{tabular}
\end{table}

\subsection*{Decision-making parameters}

We set a common level of rationality $\beta=6$ in all simulations, which captures a moderate level of rationality so that individuals tend to maximize their payoff, but always have a small but nonnegligible probability of adopting the behavior with the lower payoff. Before detailing the parameters used in the three case studies, we provide a brief discussion on the relative order of magnitude between the model parameters, which guided our choices.

The decision-making process is based on the comparison between the two payoff functions in Eqs.~(\ref{eq:pi0})--(\ref{eq:pi1}). The contribution of social influence to the payoff is always bounded between $0$ and $1$. Hence, social influence is significant if the other terms do not have a higher order of magnitude. Consequently, policy interventions $u(t)>1$ can be considered severe, since their effect is predominant with respect to social influence, while policies with $u(t)<1$ are milder. The cost of self-protective behaviors consists of two terms: the immediate cost per unit-time $c$ and the accumulation factor $\gamma$. Small values of $c$ become negligible in the decision making process, while, to avoid the immediate cost dominating the other terms, we should assume $c<1$. The accumulation factor $\gamma$ captures the cost for continued periods in which an individual adopts self-protective behaviors. To model a nonnegligible effect of the accumulation of socio-economic costs, we should guarantee that over long periods in which an individual consistently adopts self-protective behaviors,  the frustration function saturates to a value comparable to the other terms. This can be achieved by imposing that
\begin{equation}\label{eq:saturation}
    \lim_{t\to\infty}c+\sum_{s=1}^t\gamma^s c=\frac{c}{1-\gamma}\approx 1\,,
\end{equation}
yielding $c+\gamma\approx 1$ (note, the above equality was obtained using the geometric series). Specifically, values of $\gamma>1-c$ guarantees that self-protective behaviors are eventually dismissed, after the complete eradication of the disease or the policy intervention is switched off. We use the risk perception function $r(t)=kz^\alpha$ with $\alpha=1/2$ for cautious populations, $\alpha=1$ to model proportional reactions, and $\alpha=2$ for slow reacting populations. As discussed in the main article, the risk perception function determines a critical epidemic prevalence $z^*=\min\{z:r(z)>1+c\}$ that triggers the adoption of self-protective behaviors even in the absence of interventions (in the presence of accumulation, the immediate cost $c$ in the expression of $z^*$ is substituted by its saturation value from~\eqref{eq:saturation}, being $c/(1-\gamma)$). We observe  that risk perception becomes nonnegligible if $k>1+\frac{c}{1-\gamma}$. To keep consistency throughout our simulations, we set $k=3$, which is a value that verifies the condition above for all the choices of parameters $c$ and $\gamma$ we make in the simulations.

The decision-making parameters used for the two models are detailed in the following.

{\bf SIS model.} We assume that the accumulation is negligible for gonorrhea (where the use of protections has an immediate cost that typically does not accumulate, such as protective sexual barriers). Hence, for all three simulations, we fix the immediate cost to $c=0.3$ and the accumulation factor $\gamma=0$. No policy intervention is set, with $u(t) = 0$ for all $t\geq 0$. In the three simulations, we test three different risk perception functions. Specifically, we consider a cautious population with $r(z) = 3\sqrt z$ in Figs.~\ref{fig:sis1} and~\ref{fig:sis_phase1}, a population with a proportional reaction, $r(z) = 3 z$ in Figs.~\ref{fig:sis2},~\ref{fig:sis4},~\ref{fig:sis_phase2}, and~\ref{fig:sis_phase4}, and a population slow to react with $r(z)=3z^2$ in Figs.~\ref{fig:sis3} and~\ref{fig:sis_phase3}.

{\bf SIR model.} Self-protective behaviors involve social distancing and closures of economic activities, which has been shown to typically yield an accumulation of psychological distress and economic losses~\cite{nicola2020socio,Qiu2020,Bartik2020economic}. Hence, we assume  a high accumulation factor $\gamma = 0.9$ and we fix $c = 0.12$, in light of our discussion above. To capture the slow reaction of the population due to the initial suppression of information (to keep morale up during World War I)~\cite{Markel2007,Bootsma2007}, we set $r(z)=3z^2$. To further mirror real-world interventions by public authorities, in Fig.~\ref{fig:flu1}, we set an initial intervention level equal to $u(0)=0$, which switches to $u(t)=\bar u=0.5$ once $1\%$ of the population is infected and then remains active for $28$ days before being turned off again, consistent with~\cite{Markel2007}. Then, we consider three different scenarios of intervention policies. In Fig.~\ref{fig:flu2}, we set a constant mild level of interventions $u(t)=0.4$, for all $t\geq 0$.  In the other two scenarios, we set $u(0)=0$. Then, in Fig.~\ref{fig:flu3}, severe policies ($u(t)=0.7$) are implemented for $21$ days after reaching $1\%$ of infections, after which $u(t)$ is linearly reduced to $u(t)=0$ over $42$ time-steps. In the second scenario (Fig.~\ref{fig:flu4}), more severe policies ($u(t)=1.2$) are implemented for the same period of $21$ time-steps, after which $u(t)$ is linearly reduced to $u(t)=0$ over a shorter time-window of $8$ time-steps. Note that we select the intensity of policy interventions and the duration of the phased reduction  to ensure that the cumulative intervention effort, $\sum_{t} u(t)$, over the duration of a lockdown, is equal to $29.4$ in both scenarios.

\section{Simulations without social influence}

In the absence of social influence, the payoff functions reduces to
\begin{equation}\label{eq:payoff_noSI}
     \pi_i^0(t)=-u(t),\qquad
    \pi_i^1(t)=r\big(z(t)\big)-f_i(t).
\end{equation}

\begin{figure}
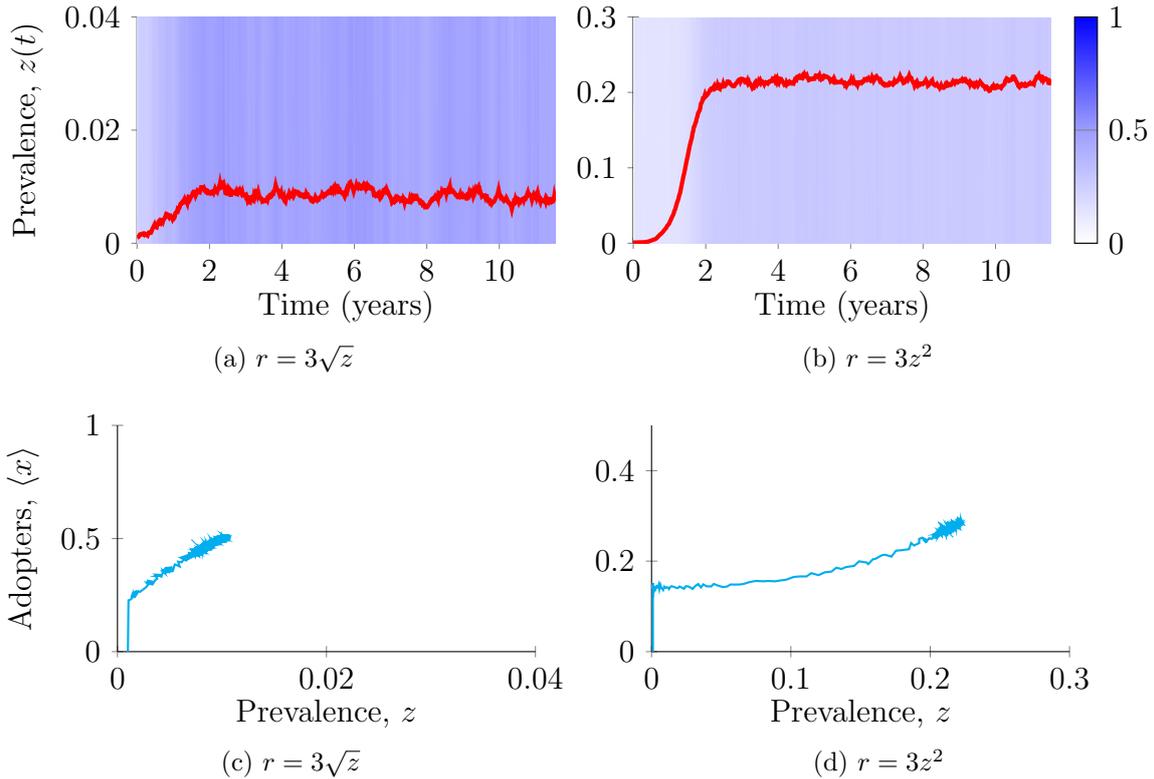

    \centering
      \subfloat[$r=3\sqrt z$]{\input{sis5.tex}\label{fig:sis5}}
      \subfloat[$r=3 z^2$]{\input{sis6.tex}\label{fig:sis6}}\\
      \subfloat[$r=3\sqrt z$]{\input{sis5_phase}\label{fig:sis_phase5}}
      \subfloat[$r=3 z^2$]{\input{sis6_phase}\label{fig:sis_phase6}\qquad}
    \caption{Simulations of the SIS model in the absence of social influence. In (a) and (b), we show the time-evolution of the  epidemic prevalence $z(t)$ (red) and the fraction of adopters of self-protections $\langle x(t)\rangle$ (intensity of the blue bands). In (c) and (d), we show the corresponding trajectories on the phase-space. Panels refer to different risk perceptions $r(z)$, as detailed in the sub-captions.    }
    \label{fig:sis_noSI}
\end{figure}

The simulations in Figs.~\ref{fig:sis4} and~\ref{fig:sis_phase4} in the main article and in Fig.~\ref{fig:sis_noSI} are obtained utilizing the payoff functions in \eqref{eq:payoff_noSI}.

\begin{thebibliography}{75}

\bibitem{Bavel2020}
Jay~J. Van~Bavel et~al.
\newblock {Using social and behavioural science to support COVID-19 pandemic
  response}.
\newblock {\em Nat. Hum. Behav.}, 4(5):460--471, May 2020.

\bibitem{Bedson2021}
Jamie Bedson, Laura~A. Skrip, Danielle Pedi, Sharon Abramowitz, Simone Carter,
  Mohamed~F. Jalloh, Sebastian Funk, Nina Gobat, Tamara Giles-Vernick, Gerardo
  Chowell, et~al.
\newblock A review and agenda for integrated disease models including social
  and behavioural factors.
\newblock {\em Nat. Hum. Behav.}, Jun 2021.

\bibitem{Pastor-Satorras2015}
R.~Pastor-Satorras, C.~Castellano, P.~Van~Mieghem, and A.~Vespignani.
\newblock Epidemic processes in complex networks.
\newblock {\em Rev. Mod. Phys.}, 87:925--979, 2015.

\bibitem{Funk2010}
S.~Funk, M.~Salath\'{e}, and V.~A. Jansen.
\newblock {Modelling the influence of human behaviour on the spread of
  infectious diseases: a review}.
\newblock {\em J. R. Soc. Interface}, 7(50):1247--1256, 2010.

\bibitem{Perra2011}
Nicola Perra, Duygu Balcan, Bruno Gonçalves, and Alessandro Vespignani.
\newblock {Towards a Characterization of Behavior-Disease Models}.
\newblock {\em PLOS One}, 6(8), 08 2011.

\bibitem{Sahneh2012}
Faryad~Darabi Sahneh, Fahmida~N. Chowdhury, and Caterina~M. Scoglio.
\newblock On the existence of a threshold for preventive behavioral responses
  to suppress epidemic spreading.
\newblock {\em Sci. Rep.}, 2(1):632, Sep 2012.

\bibitem{Granell2013}
C.~Granell, S.~G\'{o}mez, and A.~Arenas.
\newblock {Dynamical interplay between awareness and epidemic spreading in
  multiplex networks}.
\newblock {\em Phys. Rev. Lett.}, 111(12):128701, 2013.

\bibitem{RizzoPRE2014}
A.~Rizzo, M.~Frasca, and M.~Porfiri.
\newblock Effect of individual behavior on epidemic spreading in activity
  driven networks.
\newblock {\em Phys. Rev. E}, 90, 2014.

\bibitem{Wang2015}
Zhen Wang, Michael~A. Andrews, Zhi-Xi Wu, Lin Wang, and Chris~T. Bauch.
\newblock Coupled disease–behavior dynamics on complex networks: A review.
\newblock {\em Phys. Life Rev.}, 15:1 -- 29, 2015.

\bibitem{Verelst2016}
Frederik Verelst, Lander Willem, and Philippe Beutels.
\newblock {Behavioural change models for infectious disease transmission: a
  systematic review (2010-2015)}.
\newblock {\em J. R. Soc. Interface}, 13:20160820, 2016.

\bibitem{Weitz2020}
Joshua~S. Weitz, Sang~Woo Park, Ceyhun Eksin, and Jonathan Dushoff.
\newblock Awareness-driven behavior changes can shift the shape of epidemics
  away from peaks and toward plateaus, shoulders, and oscillations.
\newblock {\em Proc. Natl. Acad. Sci. USA}, 117(51):32764--32771, 2020.

\bibitem{Gozzi2021}
Nicol\`o Gozzi, Martina Scudeler, Daniela Paolotti, Andrea Baronchelli, and
  Nicola Perra.
\newblock Self-initiated behavioral change and disease resurgence on
  activity-driven networks.
\newblock {\em Phys. Rev. E}, 104:014307, Jul 2021.

\bibitem{Tuncgenc2021influence}
Bahar Tunçgenç, Marwa {El Zein}, Justin Sulik, Martha Newson, Yi~Zhao,
  Guillaume Dezecache, and Ophelia Deroy.
\newblock Social influence matters: We follow pandemic guidelines most when our
  close circle does.
\newblock {\em Br. J. Psychol.}, 112(3):763--780, 2021.

\bibitem{Poletti2010}
Piero Poletti, Bruno Caprile, Marco Ajelli, Andrea Pugliese, and Stefano
  Merler.
\newblock Spontaneous behavioural changes in response to epidemics.
\newblock {\em J. Theor. Biol.}, 260(1):31--40, 2009.

\bibitem{nicola2020socio}
Maria Nicola, Zaid Alsafi, Catrin Sohrabi, Ahmed Kerwan, Ahmed Al-Jabir,
  Christos Iosifidis, Maliha Agha, and Riaz Agha.
\newblock {The socio-economic implications of the coronavirus pandemic
  (COVID-19): A review}.
\newblock {\em Int. J. Surg.}, 78:185--193, June 2020.

\bibitem{pedro2020}
Sansao~A. Pedro, Frank~T. Ndjomatchoua, Peter Jentsch, Jean~M. Tchuenche,
  Madhur Anand, and Chris~T. Bauch.
\newblock {Conditions for a Second Wave of COVID-19 Due to Interactions Between
  Disease Dynamics and Social Processes}.
\newblock {\em Front. Phys.}, 8:428, 2020.

\bibitem{Simon2000}
Herbert~A. Simon.
\newblock {Bounded rationality in social science: Today and tomorrow}.
\newblock {\em Mind Soc.}, 1(1):25--39, Mar 2000.

\bibitem{Flaxman2020}
Seth Flaxman, Swapnil Mishra, Axel Gandy, H.~Juliette~T. Unwin, Thomas~A.
  Mellan, Helen Coupland, Charles Whittaker, Harrison Zhu, Tresnia Berah,
  Jeffrey~W. Eaton, et~al.
\newblock {Estimating the effects of non-pharmaceutical interventions on
  COVID-19 in Europe}.
\newblock {\em Nature}, 584:257--261, Jun 2020.

\bibitem{Perra2021}
Nicola Perra.
\newblock Non-pharmaceutical interventions during the covid-19 pandemic: A
  review.
\newblock {\em Phys. Rep.}, 913:1--52, 2021.

\bibitem{Piret2021}
Jocelyne Piret and Guy Boivin.
\newblock Pandemics throughout history.
\newblock {\em Front. Microbiol.}, 11:3594, 2021.

\bibitem{handbook_game}
H.~Peyton Young and Shmuel Zamir, editors.
\newblock {\em {Handbook of Game Theory with Economic Applications}}, volume~4.
\newblock Elsevier, 2015.

\bibitem{kabir2020}
K.~M.~Ariful Kabir and Jun Tanimoto.
\newblock {Evolutionary game theory modelling to represent the behavioural
  dynamics of economic shutdowns and shield immunity in the COVID-19 pandemic}.
\newblock {\em R. Soc. Open Sci.}, 7(9):201095, 2020.

\bibitem{wei2020game}
Jinyu Wei, Li~Wang, and Xin Yang.
\newblock {Game analysis on the evolution of COVID-19 epidemic under the
  prevention and control measures of the government}.
\newblock {\em PLOS ONE}, 15(10):1--16, 10 2020.

\bibitem{Just2017}
Winfried Just, Joan Salda{\~{n}}a, and Ying Xin.
\newblock Oscillations in epidemic models with spread of awareness.
\newblock {\em J. Math. Biol.}, 76(4):1027--1057, July 2017.

\bibitem{Steinegger2020}
Benjamin Steinegger, Alex Arenas, Jes\'us G\'omez-Garde\~nes, and Clara
  Granell.
\newblock Pulsating campaigns of human prophylaxis driven by risk perception
  palliate oscillations of direct contact transmitted diseases.
\newblock {\em Phys. Rev. Research}, 2:023181, May 2020.

\bibitem{Bauch2004}
Chris~T. Bauch and David J.~D. Earn.
\newblock Vaccination and the theory of games.
\newblock {\em Proc. Natl. Acad. Sci. USA}, 101(36):13391--13394, 2004.

\bibitem{Fu2010}
Feng Fu, Daniel~I. Rosenbloom, Long Wang, and Martin~A. Nowak.
\newblock Imitation dynamics of vaccination behaviour on social networks.
\newblock {\em Proc. Royal Soc. B}, 278(1702):42--49, 2011.

\bibitem{Zhang2014}
Hai-Feng Zhang, Zhi-Xi Wu, Ming Tang, and Ying-Cheng Lai.
\newblock Effects of behavioral response and vaccination policy on epidemic
  spreading --- an approach based on evolutionary-game dynamics.
\newblock {\em Sci. Rep.}, 4(1):5666, Jul 2014.

\bibitem{Zhang2017vaccination}
Hai-Feng Zhang, Pan-Pan Shu, Zhen Wang, Ming Tang, and Michael Small.
\newblock Preferential imitation can invalidate targeted subsidy policies on
  seasonal-influenza diseases.
\newblock {\em Applied Mathematics and Computation}, 294:332--342, 2017.

\bibitem{chen2019vaccine}
Xingru Chen and Feng Fu.
\newblock Imperfect vaccine and hysteresis.
\newblock {\em Proc. Royal Soc. B}, 286(1894):20182406, 2019.

\bibitem{Chang2020}
Sheryl~L. Chang, Mahendra Piraveenan, Philippa Pattison, and Mikhail
  Prokopenko.
\newblock Game theoretic modelling of infectious disease dynamics and
  intervention methods: a review.
\newblock {\em J. Biol. Dyn.}, 14(1):57--89, 2020.

\bibitem{wells2020}
Chad~R. Wells, Amit Huppert, Meagan~C. Fitzpatrick, Abhishek Pandey, Baruch
  Velan, Burton~H. Singer, Chris~T. Bauch, and Alison~P. Galvani.
\newblock {Prosocial polio vaccination in Israel}.
\newblock {\em Proc. Natl. Acad. Sci. USA}, 117(23):13138--13144, 2020.

\bibitem{brauer2011mathematical}
F.~Brauer and C.~Castillo-Chavez.
\newblock {\em Mathematical models in population biology and epidemiology}.
\newblock Springer, 2012.

\bibitem{Boccaletti2014}
S.~Boccaletti, G.~Bianconi, R.~Criado, C.I. {del Genio}, J.~Gómez-Gardeñes,
  M.~Romance, I.~Sendiña-Nadal, Z.~Wang, and M.~Zanin.
\newblock The structure and dynamics of multilayer networks.
\newblock {\em Phys. Rep.}, 544(1):1 -- 122, 2014.

\bibitem{Holme2012}
P.~Holme and J.~Saram\"{a}ki.
\newblock {Temporal networks}.
\newblock {\em Phys. Rep.}, 519:97--125, 2012.

\bibitem{Perra2012}
N.~Perra, B.~Gon\c{c}alves, R.~Pastor-Satorras, and A.~Vespignani.
\newblock {Activity driven modeling of time varying networks}.
\newblock {\em Sci. Rep.}, 2:469, 2012.

\bibitem{Holme2015}
Petter Holme.
\newblock Modern temporal network theory: a colloquium.
\newblock {\em Eur. Phys. J. B}, 88(9), 2015.

\bibitem{Valdano2015}
E.~Valdano, L.~Ferreri, C.~Poletto, and V.~Colizza.
\newblock {Analytical computation of the epidemic threshold on temporal
  networks}.
\newblock {\em Phys. Rev. X}, 5:021005, 2015.

\bibitem{Zino2016}
Lorenzo Zino, Alessandro Rizzo, and Maurizio Porfiri.
\newblock {Continuous-time discrete-distribution theory for activity-driven
  networks}.
\newblock {\em Phys. Rev. Lett.}, 117:228302, 2016.

\bibitem{Koher2019}
Andreas Koher, Hartmut H.~K. Lentz, James~P. Gleeson, and Philipp H\"ovel.
\newblock Contact-based model for epidemic spreading on temporal networks.
\newblock {\em Phys. Rev. X}, 9:031017, Aug 2019.

\bibitem{blume1995best_response}
Lawrence Blume.
\newblock {The Statistical Mechanics of Best-Response Strategy Revision}.
\newblock {\em Games Econ. Behav.}, 11(2):111--145, 1995.

\bibitem{Jackson2015}
Matthew~O. Jackson and Yves Zenou.
\newblock {Games on Networks}.
\newblock In {\em {Handbook of Game Theory with Economic Applications}},
  volume~4, chapter~3, pages 95--163. Elsevier, 2015.

\bibitem{cialdini2004social_conformity}
Robert~B Cialdini and Noah~J Goldstein.
\newblock {Social Influence: Compliance and Conformity}.
\newblock {\em Annu. Rev. Psychol.}, 55:591--621, 2004.

\bibitem{peytonyoung2015social_norms}
H~{Peyton Young}.
\newblock {The Evolution of Social Norms}.
\newblock {\em Annu. Rev. Econ.}, 7(1):359--387, 2015.

\bibitem{young1993evolution}
H~Peyton Young.
\newblock {The Evolution of Conventions}.
\newblock {\em Econometrica}, pages 57--84, 1993.

\bibitem{montanari2010spread_innovation}
Andrea Montanari and Amin Saberi.
\newblock The spread of innovations in social networks.
\newblock {\em Proc. Natl. Acad. Sci. USA}, 107(47):20196--20201, 2010.

\bibitem{young2009innovation}
H~Peyton~Young.
\newblock {Innovation Diffusion in Heterogeneous Populations: Contagion, Social
  Influence, and Social Learning}.
\newblock {\em Am. Econ. Rev.}, 99(5):1899--1924, 2009.

\bibitem{Qiu2020}
Jianyin Qiu, Bin Shen, Min Zhao, Zhen Wang, Bin Xie, and Yifeng Xu.
\newblock {A nationwide survey of psychological distress among Chinese people
  in the COVID-19 epidemic: implications and policy recommendations}.
\newblock {\em Gen. Psychiatr.}, 33(2):e100213--e100213, Mar 2020.

\bibitem{Rizzo2016Ebola}
A.~Rizzo, B.~Pedalino, and M.~Porfiri.
\newblock A network model for {Ebola} spreading.
\newblock {\em J. Theor. Biol.}, 394(7):212--222, 2016.

\bibitem{Pozzana2017}
Iacopo Pozzana, Kaiyuan Sun, and Nicola Perra.
\newblock Epidemic spreading on activity-driven networks with attractiveness.
\newblock {\em Phys. Rev. E}, 96, Oct 2017.

\bibitem{zino2018memory}
L.~Zino, A.~Rizzo, and M.~Porfiri.
\newblock Modeling memory effects in activity-driven networks.
\newblock {\em SIAM J. Appl. Dyn. Syst.}, 17(4):2830--2854, 2018.

\bibitem{Petri2018}
Giovanni Petri and Alain Barrat.
\newblock Simplicial activity driven model.
\newblock {\em Phys. Rev. Lett.}, 121:228301, 2018.

\bibitem{Leitch2019}
Jack Leitch, Kathleen~A. Alexander, and Srijan Sengupta.
\newblock Toward epidemic thresholds on temporal networks: a review and open
  questions.
\newblock {\em Appl. Netw. Sci.}, 4(1):105, 2019.

\bibitem{VanMieghem2009}
P.~{Van Mieghem}, J.~Omic, and R.~Kooij.
\newblock Virus spread in networks.
\newblock {\em IEEE/ACM Trans. Netw.}, 17(1):1--14, Feb 2009.

\bibitem{Watts1998}
Duncan~J. Watts and Steven~H. Strogatz.
\newblock Collective dynamics of `small-world' networks.
\newblock {\em Nature}, 393(6684):440--442, Jun 1998.

\bibitem{Aiello2001}
William Aiello, Fan Chung, and Linyuan Lu.
\newblock A random graph model for power law graphs.
\newblock {\em Exp. Math.}, 10(1):53--66, 2001.

\bibitem{Yorke1978}
JAMES~A. Yorke, HERBERT~W. Hethcote, and ANNETT Nold.
\newblock {Dynamics and Control of the Transmission of Gonorrhea}.
\newblock {\em Sex. Transm. Dis.}, 5(2):51--55, 1978.

\bibitem{Mills2004}
Christina~E. Mills, James~M. Robins, and Marc Lipsitch.
\newblock Transmissibility of 1918 pandemic influenza.
\newblock {\em Nature}, 432(7019):904--906, Dec 2004.

\bibitem{Markel2007}
Howard Markel, Harvey~B. Lipman, J.~Alexander Navarro, Alexandra Sloan,
  Joseph~R. Michalsen, Alexandra~Minna Stern, and Martin~S. Cetron.
\newblock {Nonpharmaceutical Interventions Implemented by US Cities During the
  1918-1919 Influenza Pandemic}.
\newblock {\em JAMA}, 298(6):644--654, 08 2007.

\bibitem{Mossong2008}
Joël Mossong, Niel Hens, Mark Jit, Philippe Beutels, Kari Auranen, Rafael
  Mikolajczyk, Marco Massari, Stefania Salmaso, Gianpaolo~Scalia Tomba, Jacco
  Wallinga, et~al.
\newblock {Social Contacts and Mixing Patterns Relevant to the Spread of
  Infectious Diseases}.
\newblock {\em PLOS Med.}, 5(3), 03 2008.

\bibitem{Barry2009}
John~M. Barry.
\newblock Pandemics: avoiding the mistakes of 1918.
\newblock {\em Nature}, 459(7245):324--325, May 2009.

\bibitem{Bartik2020economic}
Alexander~W. Bartik, Marianne Bertrand, Zoe Cullen, Edward~L. Glaeser, Michael
  Luca, and Christopher Stanton.
\newblock {The impact of COVID-19 on small business outcomes and expectations}.
\newblock {\em Proc. Natl. Acad. Sci. USA}, 117(30):17656--17666, 2020.

\bibitem{Lopez2020}
Leonardo L{\'o}pez and Xavier Rod{\'o}.
\newblock {The end of social confinement and COVID-19 re-emergence risk}.
\newblock {\em Nat. Hum. Behav.}, 4(7):746--755, Jul 2020.

\bibitem{cobey2020modeling}
Sarah Cobey.
\newblock Modeling infectious disease dynamics.
\newblock {\em Science}, 368(6492):713--714, 2020.

\bibitem{oliver2020phone}
Nuria Oliver, Bruno Lepri, Harald Sterly, Renaud Lambiotte, S{\'e}bastien
  Deletaille, Marco De~Nadai, Emmanuel Letouz{\'e}, Albert~Ali Salah, Richard
  Benjamins, Ciro Cattuto, et~al.
\newblock Mobile phone data for informing public health actions across the
  covid-19 pandemic life cycle.
\newblock {\em Sci. Adv.}, 6(23), 2020.

\bibitem{google}
{Google}.
\newblock {COVID-19 Community Mobility Reports}.
\newblock \url{https://www.google.com/covid19/mobility}, 2021.

\bibitem{Haug2020}
Nils Haug, Lukas Geyrhofer, Alessandro Londei, Elma Dervic, Am{\'e}lie
  Desvars-Larrive, Vittorio Loreto, Beate Pinior, Stefan Thurner, and Peter
  Klimek.
\newblock Ranking the effectiveness of worldwide covid-19 government
  interventions.
\newblock {\em Nat. Hum. Behav.}, 4(12):1303--1312, Dec 2020.

\bibitem{Singhe2021}
Surya Singh, Mujaheed Shaikh, Katharina Hauck, and Marisa Miraldo.
\newblock Impacts of introducing and lifting nonpharmaceutical interventions on
  covid-19 daily growth rate and compliance in the united states.
\newblock {\em Proc. Natl. Acad. Sci. USA}, 118(12), 2021.

\bibitem{Dryhurst2020}
Sarah Dryhurst, Claudia~R. Schneider, John Kerr, Alexandra L.~J. Freeman,
  Gabriel Recchia, Anne~Marthe van~der Bles, David Spiegelhalter, and Sander
  van~der Linden.
\newblock Risk perceptions of covid-19 around the world.
\newblock {\em J. Risk Res.}, 23(7-8):994--1006, 2020.

\bibitem{estrada2020covid}
Ernesto Estrada.
\newblock {COVID-19 and SARS-CoV-2. Modeling the present, looking at the
  future}.
\newblock {\em Phys. Rep.}, 869:1--51, 2020.

\bibitem{arenas2020covid}
Alex Arenas, Wesley Cota, Jes\'us G\'omez-Garde\~nes, Sergio G\'omez, Clara
  Granell, Joan~T. Matamalas, David Soriano-Pa\~nos, and Benjamin Steinegger.
\newblock Modeling the spatiotemporal epidemic spreading of covid-19 and the
  impact of mobility and social distancing interventions.
\newblock {\em Phys. Rev. X}, 10:041055, Dec 2020.

\bibitem{rugh1996linear}
W.J. Rugh.
\newblock {\em Linear System Theory}, volume~2.
\newblock Prentice Hall, Upper Saddle River, NJ, 1996.

\bibitem{Lajmanovich1976}
Ana Lajmanovich and James~A. Yorke.
\newblock A deterministic model for gonorrhea in a nonhomogeneous population.
\newblock {\em Math. Biosci.}, 28(3):221--236, 1976.

\bibitem{Prem2020}
Kiesha Prem, Yang Liu, Timothy~W. Russell, Adam~J. Kucharski, Rosalind~M. Eggo,
  Nicholas Davies, Stefan Flasche, Samuel Clifford, Carl A.~B. Pearson,
  James~D. Munday, et~al.
\newblock {The effect of control strategies to reduce social mixing on outcomes
  of the COVID-19 epidemic in Wuhan, China: a modelling study}.
\newblock {\em Lancet Public Health}, 5(5):e261--e270, May 2020.

\bibitem{Bootsma2007}
Martin C.~J. Bootsma and Neil~M. Ferguson.
\newblock {The effect of public health measures on the 1918 influenza pandemic
  in U.S. cities}.
\newblock {\em Proc. Natl. Acad. Sci. USA}, 104(18):7588--7593, 2007.

\end{thebibliography}
\end{document}